\documentclass[aps,floatfix,showpacs,preprintnumbers,amsmath,amssymb,superscriptaddress,letterpaper]{revtex4}
\usepackage{natbib}
\usepackage{dcolumn}
\usepackage{graphicx}
\usepackage{enumerate}
\usepackage{inputenc}
\usepackage{fontenc}
\usepackage{float}
\usepackage{hyperref}
\usepackage{bm,array}
\usepackage{color}
\usepackage{ulem}

\newcommand{\mnras}{Mon.~Not.~R.~Astron.~Soc.}
\newcommand{\aap}{Astron.~Astrophys.}

\newcommand{\apjs}{Astrophys.~J.~Supp.}
\newcommand{\phrd}{Phys.~Rev.~D.}
\newcommand{\jcap}{J.~Cosmol.~Astropart.~Phys.}
\newcommand{\be}{\begin{equation}}
\newcommand{\ee}{\end{equation}}
\newcommand{\bea}{\begin{eqnarray}}
\newcommand{\eea}{\end{eqnarray}}

\def\isitgr {\texttt{ISiTGR}\,}
\def\cosmomc {\texttt{CosmoMC}\,}

\def\D {\mathcal{D}}

\newcolumntype{N}{>{\centering\arraybackslash}p{1cm}}
\newcolumntype{V}{>{\centering\arraybackslash}p{2cm}}
\def\[{\begin{equation}}
\def\]{\end{equation}}
\begin{document}
\title{Constraints and tensions in testing general relativity from Planck and CFHTLenS including intrinsic alignment systematics}
\author{Jason N.  Dossett}
\email{jason.dossett@brera.inaf.it}
\affiliation{INAF -- Osservatorio Astronomico di Brera, via Emilio Bianchi
46, I-23807 Merate, Italy}
\affiliation{School of Mathematics and Physics, University of Queensland, Brisbane, QLD 4072, Australia}
\affiliation{ARC Centre of Excellence for All-sky Astrophysics (CAASTRO)}
\author{Mustapha Ishak}
\email{mishak@utdallas.edu}
\affiliation{Department of Physics, The University of Texas at Dallas, Richardson, TX 75083, USA}
\author{David Parkinson}
\email{d.parkinson@uq.edu.au}
\affiliation{School of Mathematics and Physics, University of Queensland, Brisbane, QLD 4072, Australia}
\affiliation{ARC Centre of Excellence for All-sky Astrophysics (CAASTRO)}
\author{Tamara M. Davis}
\email{tamarad@physics.uq.edu.au}
\affiliation{School of Mathematics and Physics, University of Queensland, Brisbane, QLD 4072, Australia}
\affiliation{ARC Centre of Excellence for All-sky Astrophysics (CAASTRO)}
\date{\today}
\begin{abstract}
We present constraints on testing general relativity (GR) at cosmological scales using recent data sets and assess the impact of galaxy intrinsic alignment in the CFHTLenS lensing data on those constraints. We consider data from Planck temperature anisotropies, the galaxy power spectrum from WiggleZ survey, weak lensing tomography shear-shear cross correlations from the CFHTLenS survey, Integrated Sachs Wolfe-galaxy cross correlations, and baryon acoustic oscillation data. We use three different parameterizations of modified gravity (MG), one that is binned in redshift and scale, a parameterization that evolves monotonically in scale but is binned in redshift, and a functional parameterization that evolves only in redshift. We present the results in terms of the MG parameters $Q$ and $\Sigma$. We employ an intrinsic alignment model with an amplitude $A_{\rm CFHTLenS}$ that is included in the parameter analysis. We find an improvement in the constraints on the MG parameters corresponding to $40-53\%$ increase on the figure of merit compared to previous studies, and GR is found consistent with the data at the $95\%$ confidence level. The bounds found on $A_{\rm CFHTLenS}$ are sensitive to the MG parameterization used, and the correlations between $A_{\rm CFHTLenS}$ and MG parameters are found to be weak to moderate. For all 3 MG parameterizations $A_{\rm CFHTLenS}$ is found consistent with zero when the whole lensing sample is used, however, when using the optimized early-type galaxy sample a significantly non-zero $A_{\rm CFHTLenS}$ is found for GR and the scale-independent MG parameterization. We find that the tensions observed in previous studies persist, and there is an indication that CMB data and lensing data prefer different values for MG parameters, particularly for the parameter $\Sigma$. The analysis of the confidence contours and probability distributions suggest that the bimodality found follows that of the known tension in the $\sigma_8$ parameter.
\end{abstract}
\pacs{95.36.+x,98.80.Es,98.62.Sb}
\maketitle
\section{introduction}
With the ongoing and future high precision surveys and missions such as the Dark Energy Survey (DES), Large-Synoptic Survey Telescope (LSST), Euclid and WFIRST, the question of testing general relativity at cosmological scales continues to drive a lot of interest in the quest to understand the nature of gravity and the dark energy associated with the observed cosmic acceleration. While probing gravity theories at cosmological scales is a legitimate endeavor by itself, it also tackles the question of whether the observed cosmic acceleration is due to an extension or modification to Einstein's equations at cosmological scales, or is the result of a repulsive dark energy component permeating the universe. There exists a large body of  literature on these questions. We provide a partial list references and reviews on the topic here \cite{Lue:2004,Song:2005a,Knox:2006,Ishak:2006,Linder:2005,Koivisto:2006,Koyama:2006,Zhang:2006,Zhang:2007,Hu:2007,Caldwell:2007,Kunz:2007,Huterer:2007,Linder:2007,Ishak:2007,Amendola:2008,Gabadadze:2008,Polarski:2008,Acquaviva:2008,Bertschinger:2008,Daniel:2008,Jain:2008,Wei:2008,Dent:2009,Gong:2009,Ishak:2009,Dossett:2011a,Serra:2009,Thomas:2009,Bean:2010,Daniel:2010a,Zhao:2010,BL:2011,Lombriser:2011,Toreno:2011,Song:2011,Hojjati:2011,Dossett:2011b,Hojjati:2012,Dossett:2012,Simpson:2013,Huetal:2014,Dossett:2014,Boubekeur:2014,MGreview1,MGreview2,MGreview3}, and we direct the reader to the citations therein.  

The precise results of cosmic microwave temperature anisotropies from Planck \cite{Ade:2013zuv,Ade:2013ktc,Planck:2013kta}, the recent data from the WiggleZ Dark Energy Survey\cite{Drinkwater:2010,Parkinson:2012}, and the CFHTLenS weak lensing data \cite{Erben:2013,Hildebrandt:2012,Miller:2013,Heymans:2012,Benjamin:2013} have all made it appealing to derive and analyze the latest constraints on general relativity or modified gravity parameters. These results will also help to guide us in advancing the analysis frameworks and to point us toward issues that need to be dealt with in a timely manner. 

In this paper, we contrast results from different ways to parameterize modifications to general relativity (GR).  We generically call these modified gravity (MG) parameters, and we test several different types of  parameterizations including a binned parameterization that evolves in redshift and scale, a hybrid parameterization that evolves monotonically in scale but is binned in redshift, and finally a functional parameterization that evolves only in redshift. 

In various analyses testing gravity at cosmological scales, weak lensing (cosmic shear) plays a prominent role in constraining the growth of large scale structure and thus the MG parameters. Weak lensing is also at the center of a number of future experiments aimed at testing gravity at cosmological scales (e.g. LSST, Euclid, WFIRST). However, an active area of work in the lensing community is also focused on understanding and controlling the systematic effects affecting this probe. At the forefront of these systematics are the intrinsic alignments of galaxies that generate correlations which contaminate the pure cosmic shear signal, see for example the reviews \cite{IAReview1,IAReview2} and references therein. Briefly, there exist two types of galaxy intrinsic alignments. The first one is between close galaxies aligned with each other due the gravitational field present during their formation. These are referred to as intrinsic ellipticity -- intrinsic ellipticity type or simply the II-type for the 2-point correlations and III for the 3-point correlations. The second type of intrinsic alignments are due to the fact that a massive structure aligns galaxies close to it and also produces lensing of background galaxies, resulting into an anti-correlation between cosmic shear and intrinsic ellipticities. This is known as the gravitational shear -- intrinsic ellipticity type, or the GI-type for the 2-point and can be generalized based on the same idea to the 3-point correlations giving the GGI and GII types. While the II and III intrinsic alignment can be suppressed by binning and cross-correlation techniques to assure that the galaxies are far enough and the gravitational tidal effect is small, these techniques cannot eliminate the GI, GGI and GII alignments since these are present between distant galaxies. Some theoretical analyses of the effect of intrinsic alignments and other lensing systematics on MG parameters can be found in \cite{Laszlo:2011,Kirk:2011}. We include in our analysis the possible effect of the 2-point II and GI intrinsic alignments present in the CFHTLenS data and analyze their correlations with the MG parameters in particular. 
 
There has also been increasing discussion about possible tensions between various cosmological data sets, particularly in probing the amplitude of matter fluctuations at the CMB level versus probes at lower redshifts (e.g. weak lensing, galaxy clustering), see for example \cite{MacCrann:2014} and references therein.  The CMB tends to prefer models with higher values of the extrapolated cosmological parameter $\sigma_8$ (the clustering amplitude on scales of $8h^{-1}{\rm Mpc}$) than that which is obtained from low redshift probes of the growth of structure. While some works have explored resolving these tensions with various changes to the neutrino sector (for example, \cite{Hamann:2013,Battye:2013,Battye:2014,MacCrann:2014}), it is interesting to explore here if these tensions are present or reflected on the modified gravity parameters.  

The paper is organized as follows: in section II, we describe the methodology and parameterizations used. We describe the data sets used in the analysis in section III. The results and discussion are in section IV, while we conclude in section V. 
%
%

\section{Methodology}
\label{sec:Method}
In order to constrain deviations from general relativity we update the publicly available package \isitgr \cite{Dossett:2011b,Dossett:2012,ISITGRweb} for use with the December 2013 version of \cosmomc \cite{cosmomc} which is compatible with the likelihood codes for CMB power spectrum data from the Planck satellite \cite{Ade:2013ktc}.  Here we will briefly overview the modified growth formalism used in \isitgr. A much more detailed account of the this formalism though is available in \cite{Dossett:2011b,Dossett:2012}.
\subsection{The Modified Growth Formalism of \isitgr}
\isitgr uses modified versions of the first order perturbed Einstein's equations from the perturbed Friedmann-Lema\^{i}tre-Robertson-Walker (FLRW) metric.  In a flat universe this metric is written in the conformal Newtonian gauge as:
\be
ds^2=a(\tau)^2[-(1+2\Psi)d\tau^2+(1-2\Phi)dx^idx_i],
\label{eq:FLRWNewt}
\ee
where $\tau$ is conformal time, $a(\tau)$ is the scale factor normalized to one today, the $x_i$'s are the comoving coordinates, and $\Psi$ and $\Phi$ are the potentials describing the scalar modes of the metric perturbations.

{Modified gravity parameters have been introduced in various interrelated notations that mainly parameterize a possible difference between the two potentials in the metric (for example, the gravitational slip parameter of \cite{Caldwell:2007}), and a second parameter that characterizes how the spacetime curvature side (or gravitational potentials side) is coupled to the source terms via the perturbed Einstein equations (sometimes this parameter can be related to a an effective gravitational constant). Due to degeneracies between some of these parameters, some other combined parameters have been proposed as exemplified at the end of this sub-section. We refer the reader to references \cite{Zhang:2007,Caldwell:2007,Bertschinger:2008,Daniel:2008,Jain:2008,Wei:2008,Serra:2009,Thomas:2009,Bean:2010,Daniel:2010a,Zhao:2010,Hojjati:2011,Dossett:2011a,Dossett:2011b,Hojjati:2012}. A summary of the relationships between the various parameterizations can be found in \cite{Dossett:2011a}. It is also worth mentioning that papers have used these parameterizations in a functional or binned form. We use here the notation introduced by \cite{Bean:2010} and used in \isitgr \cite{Dossett:2011b} as:}
\bea
k^2\Phi  &=& -4\pi G a^2\sum_i \rho_i \Delta_i \,  Q(k,a)
\label{eq:PoissonMod}\\
k^2(\Psi-R(k,a)\,\Phi) &=& -12 \pi G  a^2\sum_i \rho_i(1+w_i)\sigma_i \, Q(k,a).
\label{eq:Mod2ndEin}
\eea
where, with $i$ denoting a particular matter species, $\rho_i$ is the density, $\Delta_i$ is the rest-frame overdensity, and $\sigma_i$ is the shear stress. $Q(k,a)$ and $R(k,a)$ are the time and scale dependent MG parameters, which both take a value of $1$ in general relativity.  $Q$ quantifies a modification to what is often referred to as the Poisson equation (though as noted in \cite{Dossett:2014} this equation is not truly the Poisson equation as it relates the overdensity to the space-like potential, $\Phi$ which only affects relativistic particles). $R$ then represents an inherent inequality between the two potentials that may be caused by a modified gravity model, and is known as the gravitational slip \cite{Caldwell:2007}.  

In order to not only avoid a strong parameter degeneracy between the parameters $Q$ and $R$, but to also have a parameter that is directly probed by observations, \isitgr \ does not directly use Eq. \eqref{eq:Mod2ndEin} in its code, but rather uses a combination of Eqs. \eqref{eq:PoissonMod} and \eqref{eq:Mod2ndEin} \cite{Dossett:2011a}:
\be
k^2(\Psi+\Phi) = -8\pi G a^2\sum_i \rho_i \Delta_i \,\Sigma(k,a)\, -12 \pi G  a^2\sum_i \rho_i (1+w_i)\sigma_i \, Q(k,a),  
\label{eq:PoissonModSum}
\ee
where the parameter $\Sigma = Q(1+R)/2$. This parameter is directly probed by observations such as weak gravitational lensing. Like the parameters $Q$ and $R$, $\Sigma$ takes a value of $1$ in general relativity. Note that in our previous works $\Sigma$ was referred to as $\D$, but in an effort to have a consistent set of parameters in the literature going forward we are now using the much more common, $\Sigma$.

\subsection{Evolution of the MG parameters}
\label{sec:Evo}

\begin{center}
\begin{table}[t]
\begin{tabular}{|c|c|c|}\hline 
&\multicolumn{2}{|c|}{Redshift bins}\\\hline
Scale bins & $0.0<z\leq 1$ & $1 <z \leq 2$\\\hline
$0.0 < k \leq 0.01$& $Q_{1},\,\Sigma_{1}$&$Q_{3},\,\Sigma_{3}$ \\\hline
$0.01 < k< \infty$& $Q_{2},\,\Sigma_{2}$&$Q_{4},\,\Sigma_{4}$ \\\hline
\end{tabular}
\caption{\label{table:Grid}
The subscript numbering in the binned parameterizations.}
\end{table}
\end{center}

In general there are two things that can be done with this modified growth formalism. First, one can give the MG parameters a generic form in order to look for possible deviations from general relativity. Alternatively, one can assume the MG parameters take a specific functional form in order to mimic the effects of a particular modified gravity model in order to test that particular modified gravity model under the assumption that the expansion history can be described by a $w$CDM model (a model where the dark energy has an equation of state, $w$). Since our goal in this work is to look for deviations from GR, we will of course be taking the first approach.  

We use three different MG parameterizations with different time and scale dependencies. We have described these parameterizations in our previous works \cite{Dossett:2011b,Dossett:2012} and will briefly overview them again here.  
\begin{itemize}
\item{\textbf{P1}:  Firstly, we use a traditional binning parameterization in which the modified gravity parameters are binned in both redshift, $z$, and wavenumber (scale), $k$. A total of four bins are created by using two redshift bins and two scale bins. The scale bins are $k\le0.01$ and $k>0.01$, while the redshift bins are $0<z\le1$ and $1<z\le 2$.  For redshifts $z>2$ the MG parameters take their GR value of $1$ at all scales.  For continuity and numerical stability this parameterization is cast functionally as:

\be
X(k,a) =\frac{1}{2}\big(1 + X_{z_1}(k)\big)+\frac{1}{2}\big(X_{z_2}(k) - X_{z_1}(k)\big)\tanh{\frac{z-1}{0.05}}+\frac{1}{2}\big(1 - X_{z_2}(k)\big)\tanh{\frac{z-2}{0.05}},\label{eq:ZBinEvo}
\ee
with
\bea
X_{z_1}(k) &=& \frac{1}{2}\big(X_2+X_1\big)+\frac{1}{2}\big(X_2-X_1\big)\tanh{\frac{k-0.01}{0.001}},
\label{eq:kBin} \\ \nonumber
X_{z_2}(k) &=& \frac{1}{2}\big(X_4+X_3\big)+\frac{1}{2}\big(X_4-X_3\big)\tanh{\frac{k-0.01}{0.001}},
\eea
where $X$ takes the values $Q$ or $\Sigma$ so in this parameterization a total of eight MG parameters are varied, $\Sigma_i$ and $Q_i$, $i=1,2,3,4$.
}
 
\item{\textbf{P2}:  Secondly, we use a hybrid parameterization that evolves with monotonic function in scale, but is binned in redshift identical to the \textbf{P2}, Eq. \eqref{eq:ZBinEvo}.  The scale dependence of this parameterization takes the form:

\bea
X_{z_1}(k) &=& X_1 e^{-\frac{k}{0.01}}+X_2(1-e^{-\frac{k}{0.01}}), 
\label{eq:kHybridQ} \\ \nonumber
X_{z_2}(k) &=& X_3 e^{-\frac{k}{0.01}}+X_4(1-e^{-\frac{k}{0.01}}),
\eea
again giving a total of eight MG parameters, $\Sigma_i$ and $Q_i$, $i=1,2,3,4$.}

\item{\textbf{P3}:  The third and final parameterization we use is what we have previously called the functional form parameterization and was first introduced by \cite{Bean:2010}. This parameterization is scale independent and the parameters evolve only in time as:
\be
X(a) = \left(X_0-1\right)a^s +1, 
\ee
where $X$ represents one of the MG parameters and $X_0$ is the value of that parameter today.  Our use of this parameterization though is slightly different from the two we discussed above in that we evolve the MG parameters $Q$ and $R$ with this functional form rather than $Q$ and $\Sigma$.  The evolution of $\Sigma$ in this case is defined as described above using $\Sigma(a) = Q(a)(1+R(a))/2$.}
\end{itemize}
\section{Data sets}
\label{sec:datasets}
As a probe of the expansion history of the universe we use baryon acoustic oscillation (BAO) measurements from the 6dF Galaxy Survey measurement at $z= 0.106$ \cite{Beutler:2011hx}, the reanalyzed SDSS DR7 \cite{Padmanabhan:2012hf,Percival:2009xn} at effective redshift $z_{\rm eff}=0.35$, and the BOSS DR9 \cite{Anderson:2012sa} surveys at $z_{\rm eff}=0.2$ and $z_{\rm eff}=0.35$.  These data sets help to break parameter degeneracies relating to late-time observables such as $\Omega_m$.

In order to access the wealth of information contained in the CMB, we use the measurements of CMB temperature anisotropy \cite{Ade:2013zuv} from the first data release of the Planck surveyor. 
In the high-$\ell$ regime, the distribution of CMB angular power spectrum, $C_\ell$, can be well approximated by a Gaussian statistics, while the low-$\ell$ part of the $C_\ell$ distribution is non-Gaussian. For these reasons the Planck team divides the likelihood into low-$\ell$ ($\ell<50$) and high-$\ell$ ($\ell\geq 50$) parts and adopts different methodologies to build the likelihood in each region.   
The low-$\ell$ part of the likelihood employs a physically motivated Bayesian component separation technique to separate the cosmological CMB signal from diffuse Galactic foregrounds. All of the Planck frequency channels, from $30$ to $353$ GHz, are used for this part of the likelihood. 
The high-$\ell$ part of the likelihood, however, utilizes a correlated Gaussian likelihood approximation based on a fine-grained set of angular cross-spectra derived from multiple detector power-spectrum combinations between the $100$, $143$, and $217$ GHz 
frequency channels.  For this part of the likelihood, foregrounds are accounted for by marginalizing over power-spectrum foreground templates. The first data release from Planck did not include any analysis of polarization data from the satellite's observations, and for this reason the Planck likelihood code uses the low-$\ell$ WMAP polarization likelihood (WP) \cite{Ade:2013zuv,Bennett:2012zja, Komatsu:2011,Hinshaw:2013}, which is useful to break the well-known parameter degeneracy between the reionization optical depth $\tau$ and the scalar index $n_s$. Finally, unresolved foregrounds are marginalized over, assuming wide priors on the relevant nuisance parameters as described in \cite{Planck:2013kta}. 

As a first probe of large scale structure in the universe,  we use measurements of the galaxy power spectrum as made by the WiggleZ Dark Energy Survey. The WiggleZ galaxy power spectrum is measured from spectroscopic redshifts of 170,352 blue emission line galaxies over a volume of ~1 Gpc$^3$ \cite{Drinkwater:2010,Parkinson:2012}. The covariance matrices for the likelihood, as given in \cite{Parkinson:2012}, are computed using the method described by \cite{Blake:2010}.  To minimize the possible effects of non-linearities, we restrict ourselves to scales less that $k_{\rm max} = 0.2 h/{\rm Mpc}$ and use the best model proposed for non-linear corrections to the matter power spectrum for this data set as calibrated against simulations (model G in \cite{Parkinson:2012}). Finally, we also marginalize over a linear galaxy bias for each of the four redshift bins, as in \cite{Parkinson:2012}.

Our next probe of the structure of the universe is Integrated Sachs Wolfe (ISW)-galaxy cross correlations as presented in \cite{Ho:2008}.  This data set cross correlates the CMB temperature anisotropies from the ISW effect with the galaxy distributions measured from the 2MASS and SDSS luminous red galaxy (LRG) surveys.  It is a useful probe in that the galaxy distribution is sensitive to changes to the actual Poisson equation which probes changes in the parameter combination $2\Sigma-Q$, while the ISW effect directly relates to the time derivative of the sum of the metric potentials and thus probes $\Sigma$ and $\dot{\Sigma}$.  This likelihood and its usefulness in testing for deviations from general relativity has been discussed extensively in previous works, for example, \cite{Bean:2010,Zhao:2010,Daniel:2010a,Dossett:2011a,Dossett:2011b,Dossett:2012}. 

As a final probe of the matter distribution of the universe we use the weak lensing tomography shear-shear cross correlation data.  This data is not only useful in that it helps to constrain both the expansion history and the growth history of structure in the universe, but, as discussed above, it also gives us a way to directly probe the value of the MG parameter $\Sigma$. In this work we use tomographic shear cross-correlation data from the Canada France Hawaii Telescope Lensing Survey (CFHTLenS) \cite{Heymans:2013}.  The CFHTLenS survey analysis combined weak lensing data processing with THELI \cite{Erben:2013}, shear measurement with lensfit \cite{Miller:2013}, and photometric redshift measurement with PSF-matched photometry \cite{Hildebrandt:2012}. Analysis of systematic and photometric redshift errors for the weak lensing data was discussed in \cite{Heymans:2012,Benjamin:2013}. The data set of \cite{Heymans:2013} consists of 21 sets of cosmic shear correlation functions associated with six redshift bins, each spanning the angular range of $1.5\, <\, \theta\, <\, 35$ arcmin. The calculation of the theoretical shear cross correlation functions for this data set differ somewhat from those used in the weak lensing data set previously included in \isitgr, \cite{Schrabback:2010}, so we will briefly overview the relevant equations for the calculation here.  A more detailed description, though, can be found in \cite{Heymans:2013}.

As usual, the shear cross correlation functions $ \xi^{kl}_{+,-}(\theta)_{_{\rm GG}}$ between bins $k,l$ are given by
\be 
\xi^{kl}_{+,-}(\theta)_{_{\rm GG}} = \frac{1}{2\pi} \int^{\infty}_{0} d\ell\ \ell\  J_{0,4}(\ell\theta)P^{kl}_{\kappa} (\ell),
 \label{eq:ShearCrossCorrelations}
\ee
where $J_n$ is the $n^{th}$-order Bessel function of the first kind, $\ell$ is the modulus of the two-dimensional wave vector, and $P^{kl}_{\kappa}$ is the convergence cross-power spectra between bins $k$ and $l$ is given by \cite{Kaiser:1998}
\be
P_\kappa^{kl}(\ell) = \int^{\chi_h}_{0}  d\chi\, g_k(\chi)g_l(\chi)\, P_{\phi,\phi} \Big(\frac{\ell}{f_K(\chi)},\chi \Big),
\label{eq:LCPS}
\ee
with comoving radial distance, $\chi$, comoving distance to the particle horizon, $\chi_h$, and comoving angular diameter distance, $f_K(\chi)$.  Above, we have absorbed the usual extra terms into the power spectrum of the sum of the metric potentials, $P_{\phi,\phi}$, where $\phi \equiv \frac{\Phi+\Psi}{2}$, including the MG parameters. The weighted geometric lens-efficiency factor for the $k^{th}$ bin, $g_k(\chi)$, is given by 
\be 
g_k(\chi)  \equiv \frac{1}{a(\chi)}\int^{\chi_h}_{\chi} d\chi' p_k(\chi') \frac{f_K(\chi'-\chi)}{f_K(\chi')},
\label{eq:LensEff}
\ee
corresponding to the normalized galaxy redshift distributions $p_k$.

In order to mediate the effects of possible intrinsic alignment contamination to the weak lensing signal, we follow the technique used in \cite{Heymans:2013} which parameterizes the contribution of the intrinsic alignments to the shear correlation measurements using a nonlinear intrinsic alignment model introduced by \cite{Bridle:2007}, which is based on the linear tidal field alignment model of \cite{HirataSeljak:2004}, which in turn is based on earlier work of \cite{Catelan:2001}. 

 In this technique the measured correlation functions are considered to be a sum of contributions from intrinsic alignments, which we will denote $\xi^{kl}_{+,-}(\theta)_{_{\rm GI}}$ and $\xi^{kl}_{+,-}(\theta)_{_{\rm II}}$, and the true shear correlation function, $\xi^{kl}_{+,-}(\theta)_{_{\rm GG}}$ above,
\be
\hat{\xi}^{kl}_{+,-}(\theta)=\xi^{kl}_{+,-}(\theta)_{_{\rm II}}\,+\,\xi^{kl}_{+,-}(\theta)_{_{\rm GI}}\,+\,\xi^{kl}_{+,-}(\theta)_{_{\rm GG}}.
\label{eq:TheoryCorrelations}
\ee
The intrinsic alignment contributions, $\xi^{kl}_{+,-}(\theta)_{_{\rm GI}}$ and $\xi^{kl}_{+,-}(\theta)_{_{\rm II}}$, are calculated in the same way as $\xi^{kl}_{+,-}(\theta)_{_{\rm GG}}$, Eq. \eqref{eq:ShearCrossCorrelations} except that the convergence cross power spectrum, $P_\kappa$, is  replaced by the projected GI and II power spectrum, $P_{_{\rm GI}}$ and $P_{_{\rm II}}$ respectively.  Explicitly these power spectra are given by:

\be
P_{_{\rm GI}}^{kl}(\ell) = \int^{\chi_h}_{0}  d\chi\, \frac{g_k(\chi)p_l(\chi)+g_l(\chi)p_k(\chi)}{f_K(\chi)}\, F_{\rm I}\, P_{\phi,\delta_0} \Big(\frac{\ell}{f_K(\chi)},\chi \Big),
\label{eq:GICPS}
\ee

\be
P_{_{\rm II}}^{kl}(\ell) = \int^{\chi_h}_{0}  d\chi\, \frac{p_k(\chi)p_l(\chi)}{\left[f_K(\chi)\right]^2}\, F_{\rm I}^2\, P_{\delta_0,\delta_0} \Big(\frac{\ell}{f_K(\chi)},\chi \Big),
\label{eq:IICPS}
\ee
where $\delta_0$ is the matter overdensity today and $F_{\rm I}$ is a cosmology dependent factor given by:
\be
F_{\rm I} = -A_{\rm CFHTLenS}\, C_1\, \rho_{\rm crit}\, \Omega_m.
\label{eq:FCFHT}
\ee
Above, $\rho_{\rm crit}$ is the critical density of the universe today, $C_1$ is a constant with a value $5\times 10^{-14}h^{-2}M_{\odot}^{-1}{\rm Mpc}^3$, and $A_{\rm CFHTLenS}$ is a nuisance parameter that we will marginalize over in our likelihood analysis. 

\section{Results and Analysis}
\label{sec:Res}

For all results we fit for the MG parameters and relevant nuisance parameters for the various data sets, as well as the six core cosmological parameters: $\Omega_bh^2$ and $\Omega_c h^2$, the baryon and cold dark matter physical density parameters, respectively; $\theta$, the ratio of the sound horizon to the angular diameter distance of the surface of last scattering; $\tau$, the reionization optical depth; $n_s$, the spectral index; and $\ln10^{10} A_s$, the amplitude of the primordial power spectrum.  As this analysis aims to test general relativity, primarily the $\Lambda$CDM model, we assume a $\Lambda$CDM expansion history throughout our analysis.  This assumption should have little impact on our results.  As shown in \cite{Dossett:2013} the MG parameters are robust to deviations from this expansion history, even those which include dark energy with perturbations.

\begin{figure}
\centering
\begin{tabular}{cccc}
{\includegraphics[width=1.65in,height=1.65in,angle=0]{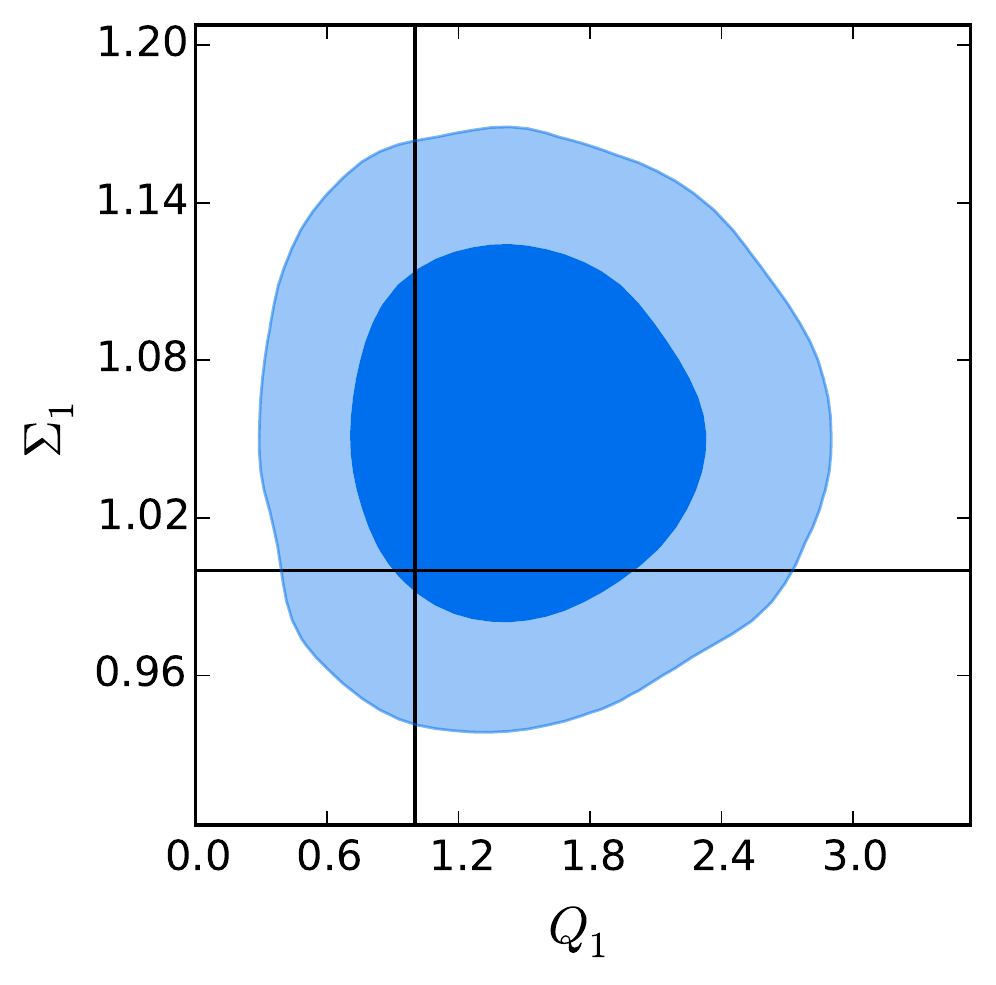}} &
{\includegraphics[width=1.65in,height=1.65in,angle=0]{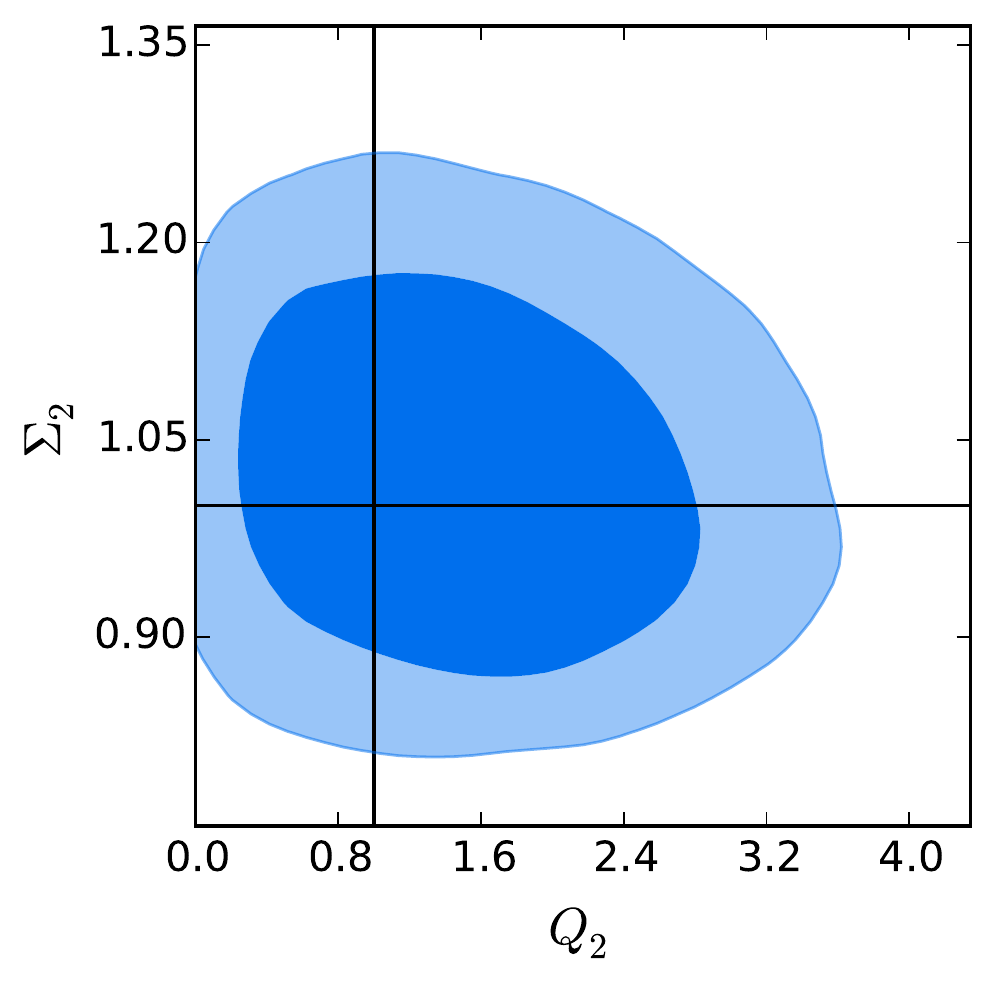}}& 
{\includegraphics[width=1.65in,height=1.65in,angle=0]{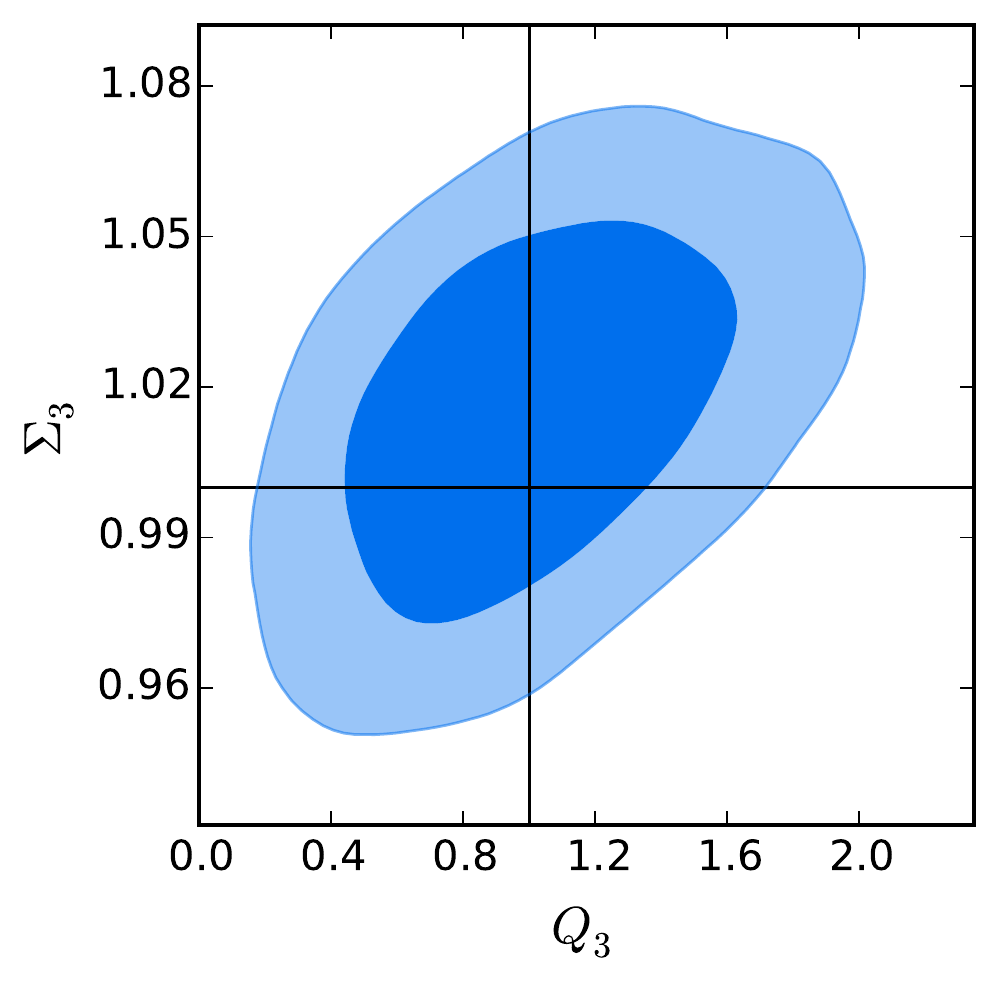}} &
{\includegraphics[width=1.65in,height=1.65in,angle=0]{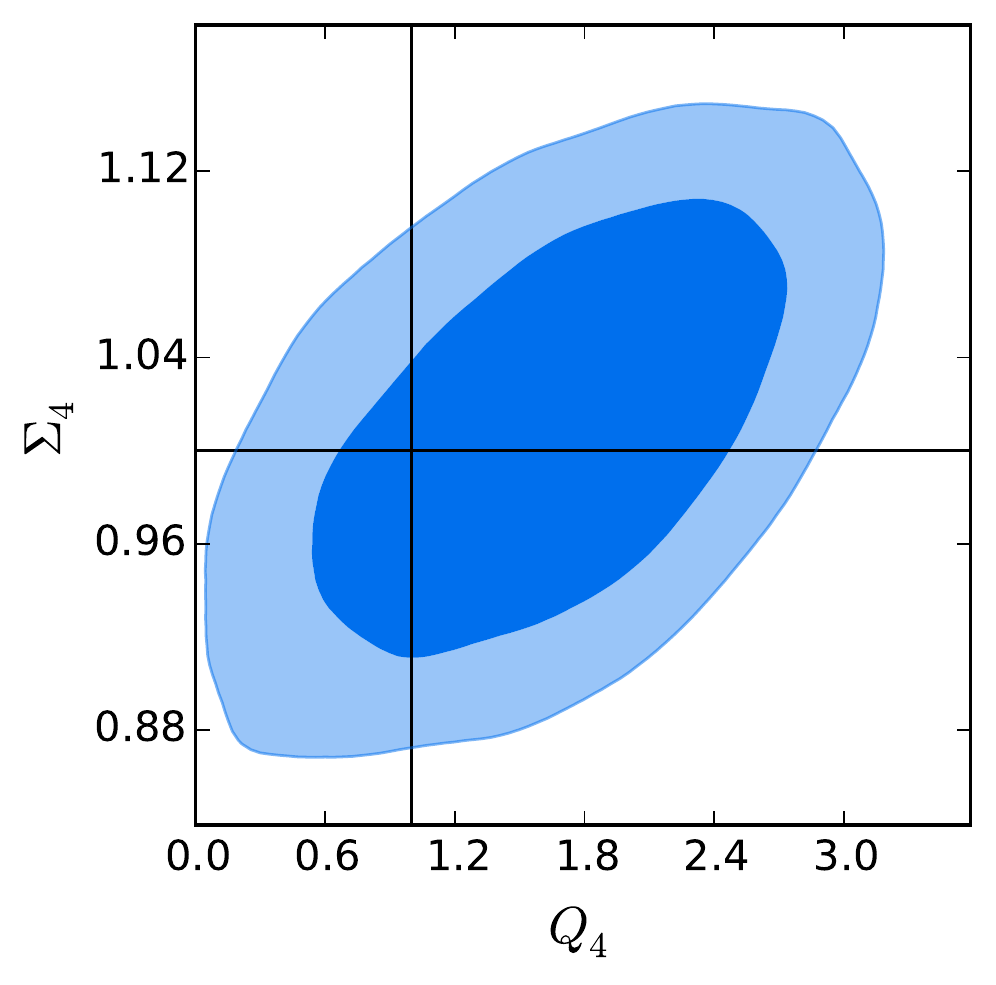}} 
\end{tabular}
\caption{\label{fig:BIN}
$68\%$ and $95\%$ 2-D confidence contours for the parameters $Q_{i}$ and $\Sigma_{i}$ from parameterization \textbf{P1} for redshift and scale dependence of the MG parameters. All of the constraints for this evolution method are fully consistent with GR at the $68\%$ level.} 
\end{figure}

\begin{table}[th]
\centering
\begin{tabular}{|N|V|N|V|}\hline
\multicolumn{4}{|c|}{{95\% confidence limits on MG parameters}}\\
\multicolumn{4}{|c|}{{evolved using form \textbf{P1}}}\\ \hline
${\bf Q_1}$&[0.49,2.56]&${\bf \Sigma_1}$&[0.97,1.14]\\ \hline
${\bf Q_2}$&[0.05,3.08]&${\bf \Sigma_2}$&[0.84,1.22]\\ \hline
${\bf Q_3}$&[0.30,1.78]&${\bf \Sigma_3}$&[0.97,1.06]\\ \hline
${\bf Q_4}$&[0.28,2.88]&${\bf \Sigma_4}$&[0.90,1.12]\\ \hline
\end{tabular}
\caption{\label{table:BinCon}We list the $95\%$ confidence limits for the MG parameters from using form \textbf{P1} to define their time and scale dependence. In this traditional binning approach we find that all of the MG parameters are fully consistent with their GR values of $1$ at the $95\%$ level.}
\end{table}

\subsection{\textbf{P1} Traditional Binning}
\label{sec:ResBin}
We first, presents our results when using form \textbf{P1} for the evolution of the MG parameters. The one dimensional marginalized constraints for this parameterization are presented in Table \ref{table:BinCon}.  We also show the 2-D marginalized $68\%$ and $95\%$ confidence contours for the parameters in Fig. \ref{fig:BIN}. Using this evolution method, all of the MG parameters are consistent with their GR value of $1$ at the $95\%$ level and no noticeable tensions are evident. This is in contrast to the other evolution methods we use where some noticeable tensions appear, making those results more revealing as we will discuss more in the sections below.

\subsection{\textbf{P2} Hybrid Evolution}
\label{sec:ResHybrid}

\begin{figure}
\centering
\begin{tabular}{cccc}
{\includegraphics[width=1.65in,height=1.65in,angle=0]{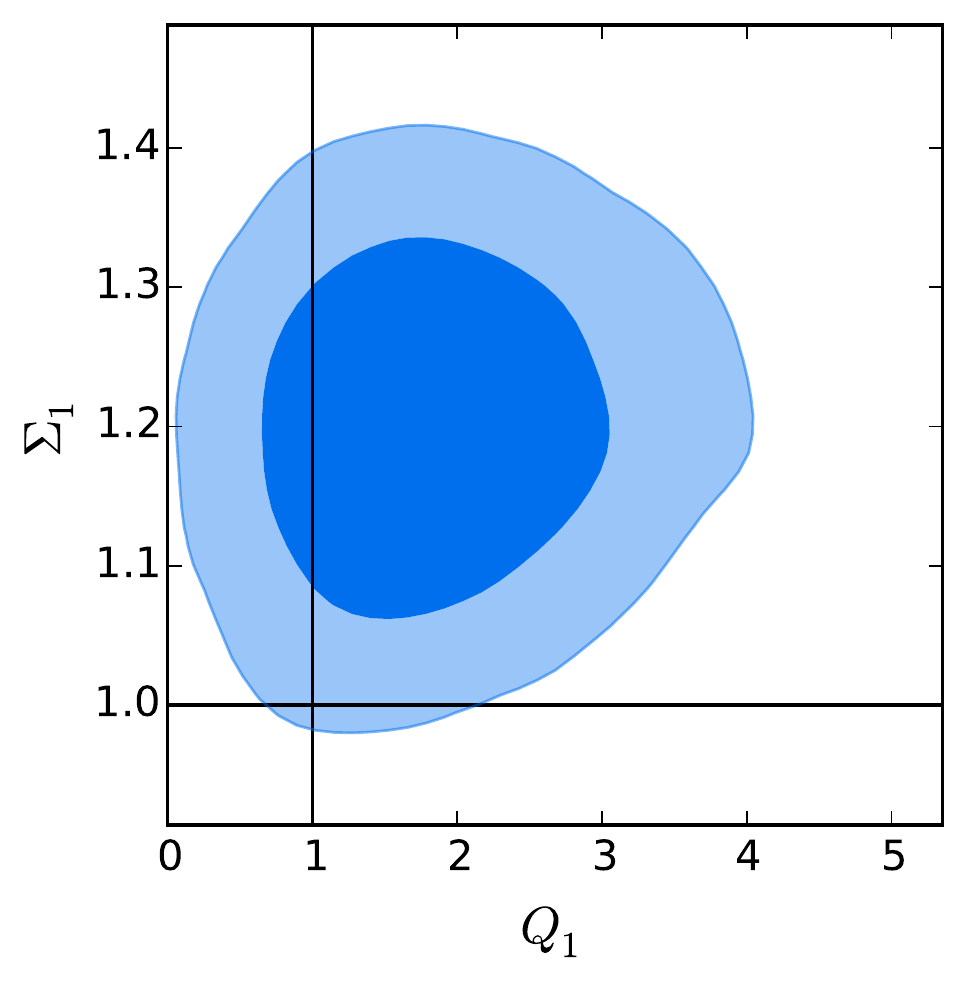}} &
{\includegraphics[width=1.65in,height=1.65in,angle=0]{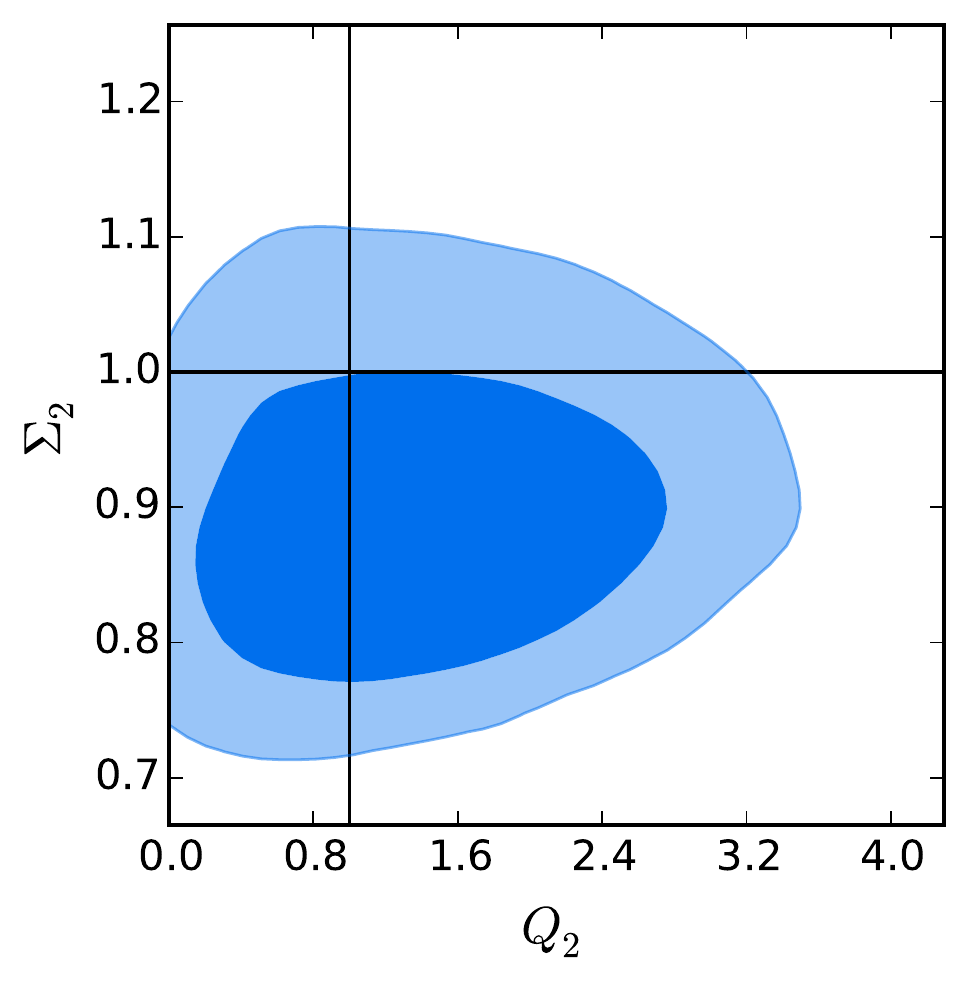}}& 
{\includegraphics[width=1.65in,height=1.65in,angle=0]{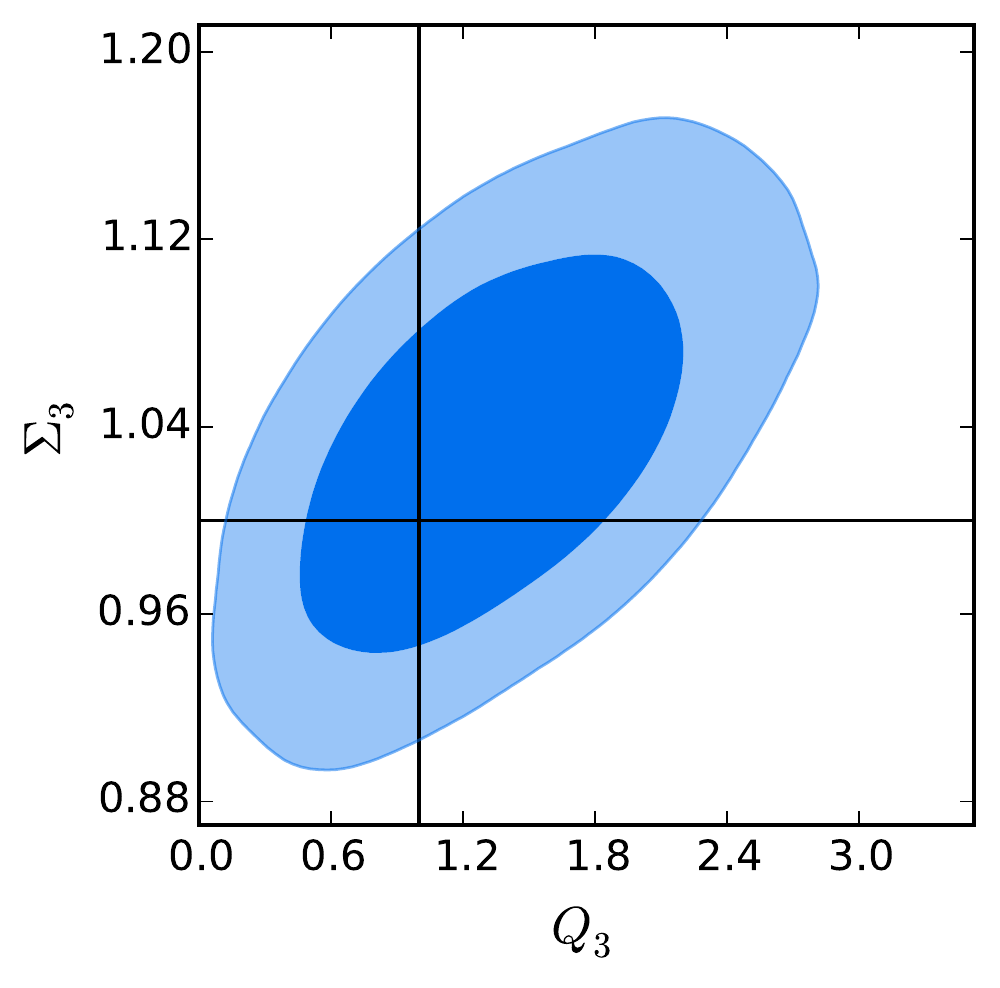}} &
{\includegraphics[width=1.65in,height=1.65in,angle=0]{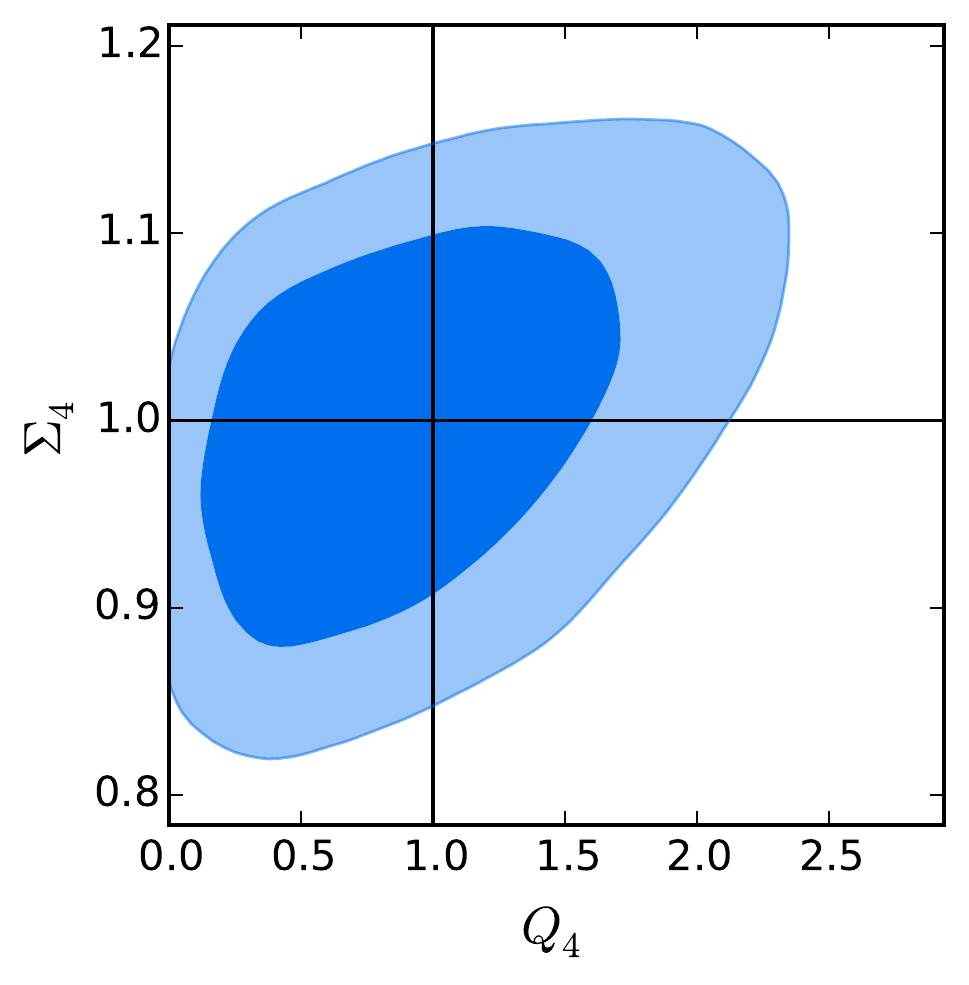}} 
\end{tabular}
\caption{\label{fig:HYBRID}
$68\%$ and $95\%$ 2-D confidence contours for the parameters $Q_{i}$ and $\Sigma_{i}$ from parameterization \textbf{P2} for redshift and scale dependence of the MG parameters. As you can see in the first bin, there a tension with the GR value of $1$. However, contrary to the marginalized 1-D constraints given in Table \ref{table:HybridCon} the GR point is still within the $95\%$ confidence region.} 
\end{figure}

The results when using the hybrid evolution method for the MG parameters, \textbf{P2}, are more interesting.  This can be seen quickly in Table \ref{table:HybridCon} where we give the marginalized constraints on the MG parameters for this evolution method.  Of particular interest are the constraints on the parameter $\Sigma_1$ where the GR value of 1 lies outside of the $95\%$ confidence interval.  This is very interesting as it could signal a possible deviation from general relativity.  Looking at the 2-D confidence contours for the MG parameters in Fig. \ref{fig:HYBRID} offers a bit of a reprieve though as the GR point still lies within the $95\%$ confidence level in the 2-D parameter space.  While it is well known that when using a large number of parameters, one of them may seem to indicate new physics when there is none (the so-called ``look-elsewhere effect''),  the tension exhibited in the constraints on this parameter, $\Sigma_1$, which again is only one of eight MG parameters, should nevertheless be explored.  We delay the majority of this discussion to \S \ref{sec:Tension}, where after presenting the results from evolution method \textbf{P3} it becomes more clear that this constraint is driven by a long standing tension between the weak lensing and CMB data sets.

\begin{table}[ht!]
\centering
\begin{tabular}{|N|V|N|V|}\hline
\multicolumn{4}{|c|}{{95\% confidence limits on MG parameters}}\\
\multicolumn{4}{|c|}{{evolved using form \textbf{P2}}}\\ \hline
${\bf Q_1}$&[0.38,3.43]&${\bf \Sigma_1}$&[1.03,1.37]\\ \hline
${\bf Q_2}$&[0.00,2.86]&${\bf \Sigma_2}$&[0.75,1.07]\\ \hline
${\bf Q_3}$&[0.28,2.46]&${\bf \Sigma_3}$&[0.93,1.14]\\ \hline
${\bf Q_4}$&[0.05,1.99]&${\bf \Sigma_4}$&[0.86,1.14]\\ \hline
\end{tabular}
\caption{\label{table:HybridCon}
We list the $95\%$ confidence limits for the MG parameters from using form \textbf{P2} to define their time and scale dependence. While most of the MG parameters for this hybrid evolution method are consistent with their GR values of $1$ at the $95\%$ level, we find a significant tension for the MG parameter $\Sigma$ in the large scale (small $k$) low redshift bin ($\Sigma_1$).}
\end{table}

\begin{center}																	
\begin{table}[ht!]																
\begin{tabular}{|V|c|c|c|c|c|c|c|c|}\hline 									
\multicolumn{9}{|c|}{Correlation table}\\							
\multicolumn{9}{|c|}{{Binning parameterization (\textbf{P1})}}\\							%
\hline																	
	&	$Q_1$	&	$Q_2$ 	&	$Q_3$	&	$Q_4$	&	$\Sigma_1$	&	$\Sigma_2$ 	&	$\Sigma_3$ 	&	$\Sigma_4$	\\ \hline
$A_{\rm CFHTLenS}$	&	-0.021162	&	-0.29209	&	0.015916	&	0.0056355	&	-0.0014863	&	0.083586	&	0.015755	&	0.066954	\\ \hline
$\sigma_8$	&	-0.012168	&	-0.53048	&	0.044293	&	-0.43088	&	0.045781	&	-0.61952	&	0.048845	&	-0.29894	\\ \hline
$\Omega_m$	&	-0.0012586	&	-0.072645	&	-0.051569	&	0.11762	&	-0.08057	&	-0.085185	&	-0.033916	&	-0.17292	\\ \hline
\multicolumn{9}{|c|}{{Hybrid parameterization (\textbf{P2})}}\\							\hline																	
$A_{\rm CFHTLenS}$	&	0.058535	&	-0.29535	&	-0.052588	&	0.095984	&	-0.14858	&	0.20636	&	-0.086038	&	0.10421	\\ \hline
$\sigma_8$	&	0.2655	&	-0.70809	&	0.12172	&	-0.33026	&	0.32713	&	-0.59009	&	0.1362	&	-0.20504	\\ \hline
$\Omega_m$	&	0.027229	&	-0.065934	&	-0.028016	&	0.0803	&	0.01565	&	-0.15645	&	0.14932	&	-0.26513	\\ \hline
\end{tabular}																	
\caption{\label{table:IA1}Correlations between $A_{\rm CFHTLenS}$, $\sigma_8$, $\Omega_m$ versus the MG parameters of \textbf{P1} and \textbf{P2}.}										
\end{table}																	
\end{center}																	
\subsection{\textbf{P3} Scale Independent Evolution}
\label{sec:ResFunc}
The results from the scale independent evolution method, \textbf{P3}, are also much more revealing than those from \textbf{P1}.  They offer some insight as to the origin of the noticeable tensions exhibited by parameters from \textbf{P3}.  The marginalized $95\%$ confidence limits for the parameters $Q_0$, $\Sigma_0$ and $R_0$ are presented in Table \ref{table:FuncCon}. The constraints are in general consistent with those of \cite{Simpson:2013}, although a direct comparison is not very simple due to different additional probes used in the two works beside the CFHTLenS data. Now, looking only at these constraints, one might conclude that nothing interesting is happening in this parameterization of the MG parameters as all of the parameters are completely consistent with one at the $95\%$ level.

\begin{table}[ht!]
\centering
\begin{tabular}{|N|V|N|V|N|V|}\hline
\multicolumn{6}{|c|}{{95\% confidence limits on MG parameters}}\\
\multicolumn{6}{|c|}{{evolved using form \textbf{P3}}}\\ \hline
${\bf Q_0}$&[0.77,1.99]&${\bf \Sigma_0}$&[0.79,1.16]&${\bf R_0}$&[-0.23,1.18]\\ \hline
\end{tabular}
\caption{\label{table:FuncCon}
We list the $95\%$ confidence limits for the parameters $Q_0$, $\Sigma_0$, and $R_0$ from the scale independent method, \textbf{P3}, of defining the evolution of the MG parameters. These constraints show no apparent tensions with GR.  This is in contrast to the noticeable tensions that become evident when looking at the 2-D confidence contours shown in Fig. \ref{fig:FUNC2D}.  The reason for this discrepancy is can be explained by looking at the 1-D probability distributions in Fig. \ref{fig:FUNC1D} where the non-Gaussianity of the parameter constraints is easily seen. 
}
\end{table}

\begin{figure}[ht!]
\centering
\begin{tabular}{cc}
{\includegraphics[width=1.65in,height=1.65in,angle=0]{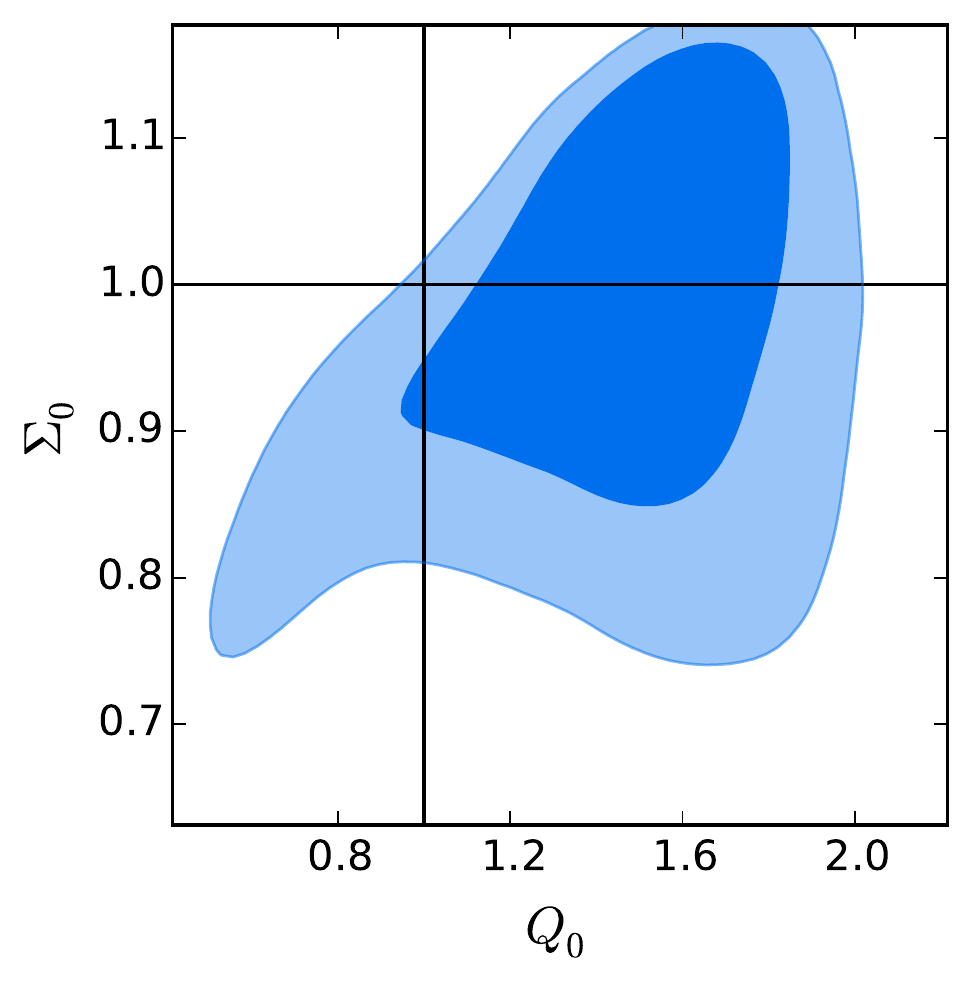}} &
{\includegraphics[width=1.65in,height=1.65in,angle=0]{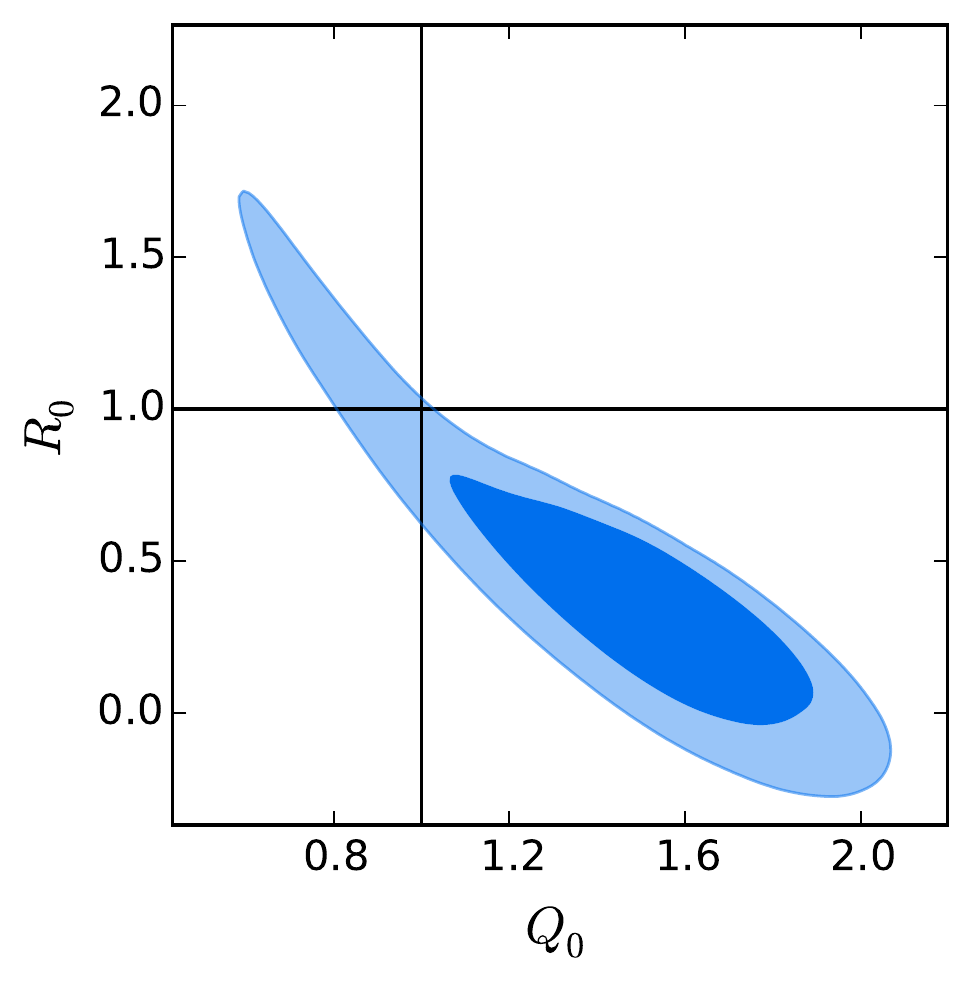}}
\end{tabular}
\caption{\label{fig:FUNC2D}
$68\%$ and $95\%$ 2-D confidence contours for the parameters $Q_{0}$, $\Sigma_0$, and $R_0$ from the scale independent parameterization, \textbf{P3}, for the MG parameters. These constraints are consistent with GR a the $95\%$ level, but a tension is evident. The tension is evident when viewing these plots is not easily seen using the 1-D constraints given in Table \ref{table:FuncCon}.  This is due to the non-Gaussianity of the probability distribution for these parameters as further seen in Fig. \ref{fig:FUNC1D}.} 
\end{figure}

\begin{figure}[t]
\centering
\begin{tabular}{ccc}
{\includegraphics[width=2.2in,height=1.65in,angle=0]{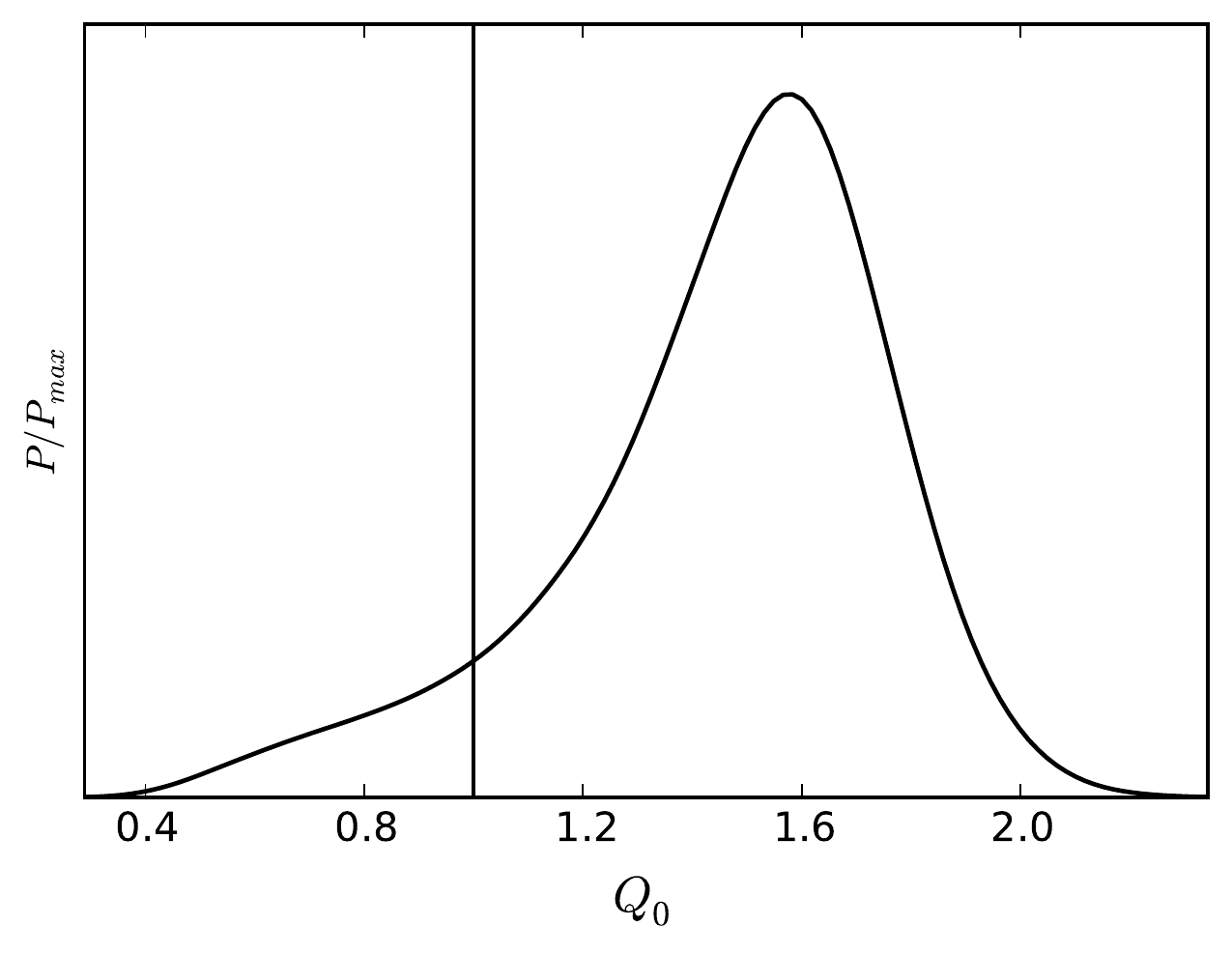}} &
{\includegraphics[width=2.2in,height=1.65in,angle=0]{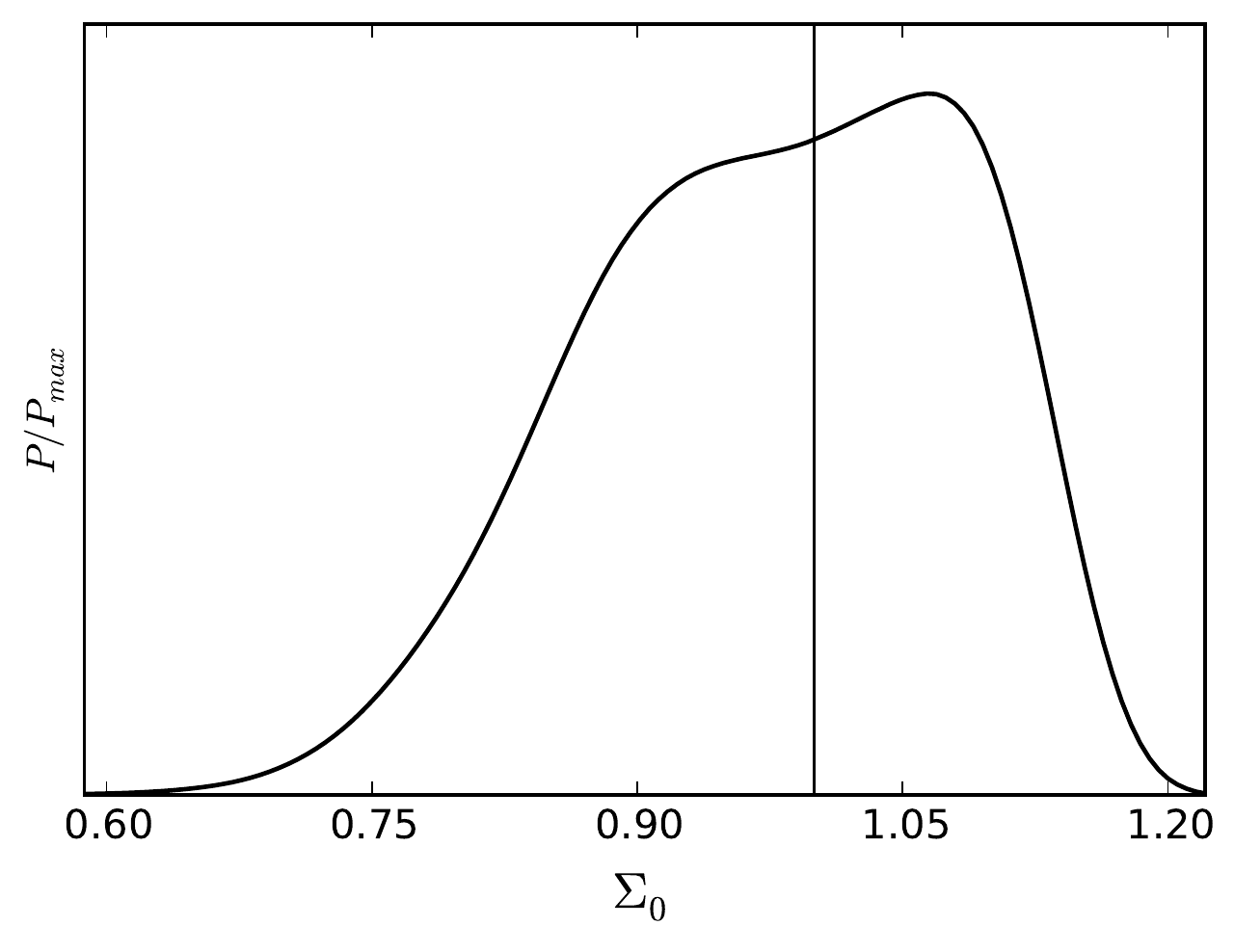}} &
{\includegraphics[width=2.2in,height=1.65in,angle=0]{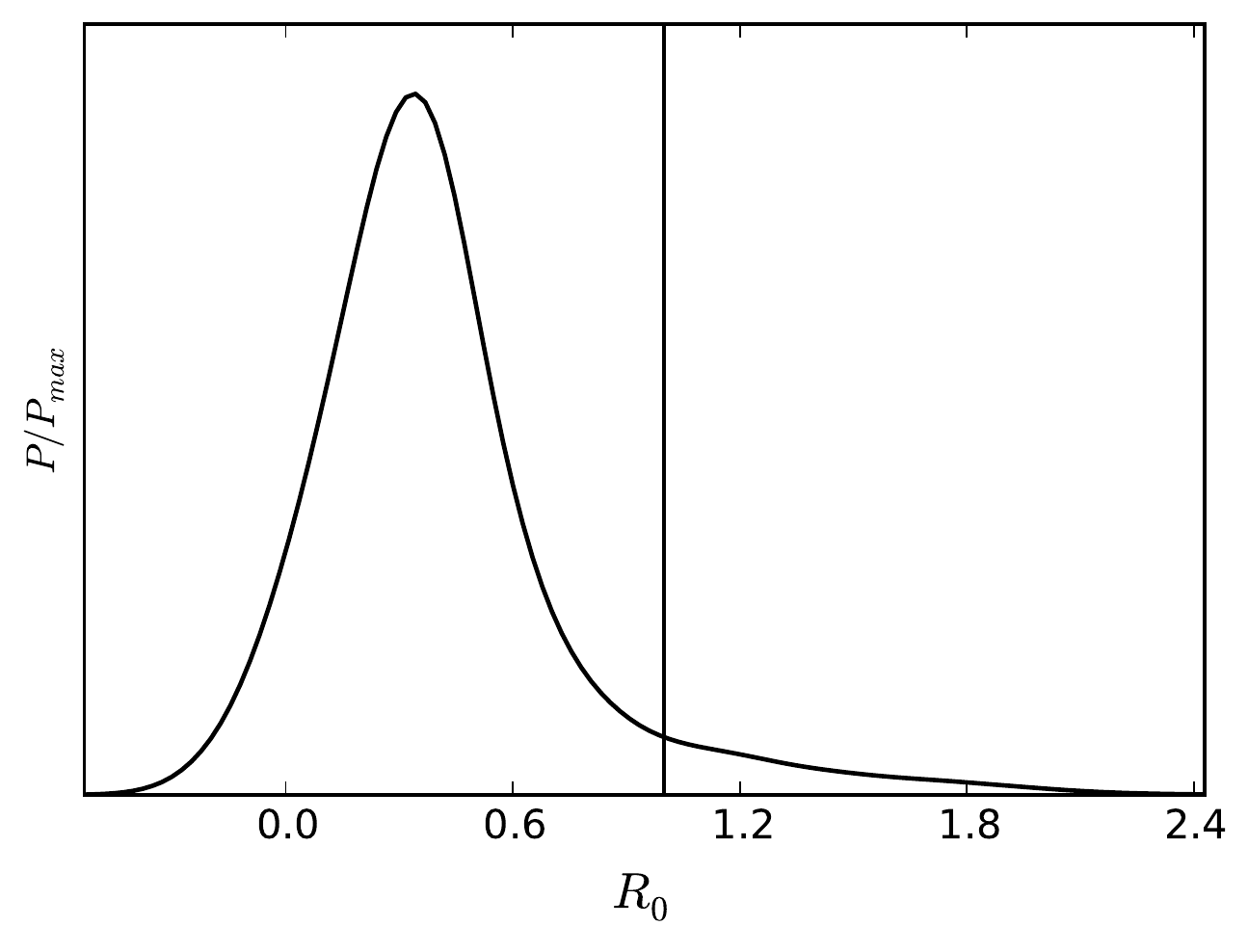}}
\end{tabular}
\caption{\label{fig:FUNC1D}
1-D probability distributions for the parameters $Q_{0}$, $\Sigma_0$, and $R_{0}$ from the scale independent parameterization for the MG parameters, \textbf{P3}. These constraints are consistent with GR a the $95\%$ level, but a tension is evident particularly in the parameters $Q_0$ and $R_0$.  The probability distributions display a significant level of non-Gaussianity, especially $\Sigma_0$ which is somewhat bimodal. This non-Gaussianity explains why the tension seen in these plots is not seen using the marginalized constraints given in Table \ref{table:FuncCon}.} 
\end{figure}

The 2-D confidence contours for these parameters plotted in Fig. \ref{fig:FUNC2D}, however, tell a very different story.  Looking at these plots it becomes apparent that there is a noticeable tension in the parameter space, with the parameters showing some preference for non-GR values.  The constraints are also non-Gaussian.  The non-Gaussianity of the parameter constraints is illustrated even better in Fig. \ref{fig:FUNC1D} where we plot the 1-D probability distributions for the MG parameters used in this evolution method.  The distributions for the parameters $Q_0$ and $R_0$ are both quite skewed, with a large tail in one end of each of the distributions, while distribution for $\Sigma_0$ is almost bimodal.  As discussed briefly in \S \ref{sec:ResHybrid} the cause of these tensions in the MG parameter space seems to be a tension between the CMB and weak lensing data sets as can be seen in figure \ref{fig:FUNCs8}.  We discuss this in depth in the next section.  
\begin{center}												
\begin{table}[t]												
\begin{tabular}{|V|c|c|}\hline 			
\multicolumn{3}{|c|}{Correlation table}\\									
\multicolumn{3}{|c|}{{Functional parameterization (\textbf{P3})}}\\					
\hline	
	&	$Q_0$	&	$\Sigma_0$	\\ \hline							
$A_{\rm CFHTLenS}$	&	-0.023164	&	0.10624	\\ \hline							
$\sigma_8$	&	-0.66775	&	-0.75738	\\ \hline							
$\Omega_m$	&	-0.072171	&	0.052317	\\ \hline							
\end{tabular}												
\caption{\label{table:IA2}Correlations between $A_{\rm CFHTLenS}$, $\sigma_8$, $\Omega_m$ versus the MG parameters of \textbf{P3}.}									
\end{table}												
\end{center}												
\subsection{Tensions Between the CMB and Weak Lensing Data Sets}
\label{sec:Tension}

As we have discussed above, there are noticeable tensions with GR in the constraints on the MG parameters from evolution methods \textbf{P2} and \textbf{P3}. One of the eight parameters from \textbf{P2} even shows an apparent deviation from GR at the $95\%$ level when looking at the marginalized confidence limits. We have argued briefly above that this tension with GR in the MG parameter constraints arises from a tension between the CMB (Planck) and weak-lensing (CFHTLenS) data sets.  We present a more in depth argument for that point here.

There has been a known tension between CMB and weak-lensing data sets for quite some time (see \cite{MacCrann:2014} and references therein).  During an analysis of the data using the $\Lambda$CDM model, this tension is usually evident from comparing the preferred $\sigma_8$ values of each of the data sets.  While the CMB data usually prefers a $\sigma_8 \sim 0.83$ \cite{Ade:2013zuv}, lensing data usually prefers a much lower $\sigma_8 \sim 0.7$.   Recently there has been a lot of work on resolving this tension between these data sets by allowing for a sterile neutrino species or heavier set of active neutrinos \cite{Hamann:2013,Battye:2013,Battye:2014,MacCrann:2014}.  In principle this would resolve the tension, as the neutrinos would suppress in late time growth and therefore explain why late time measurements of $\sigma_8$ from lensing do not match those from the CMB, which infers the value of $\sigma_8$ from early universe observations rather than directly measuring it.

The biggest indication that the tension in the MG parameter space is coming from the known tension between the two data sets is the bimodal distribution of $\Sigma_0$ from \textbf{P3}.  This of course indicates that two very different values of $\Sigma_0$ are equally preferred by the overall combination of data sets.  Since $\sigma_8$ has been useful to illustrate the tension between these two data sets before, we explore the values preferred when using \textbf{P3}.  We find very interesting results upon doing this.  

\begin{figure}[h!]
\centering
\begin{tabular}{ccc}
{\includegraphics[width=1.65in,height=1.65in,angle=0]{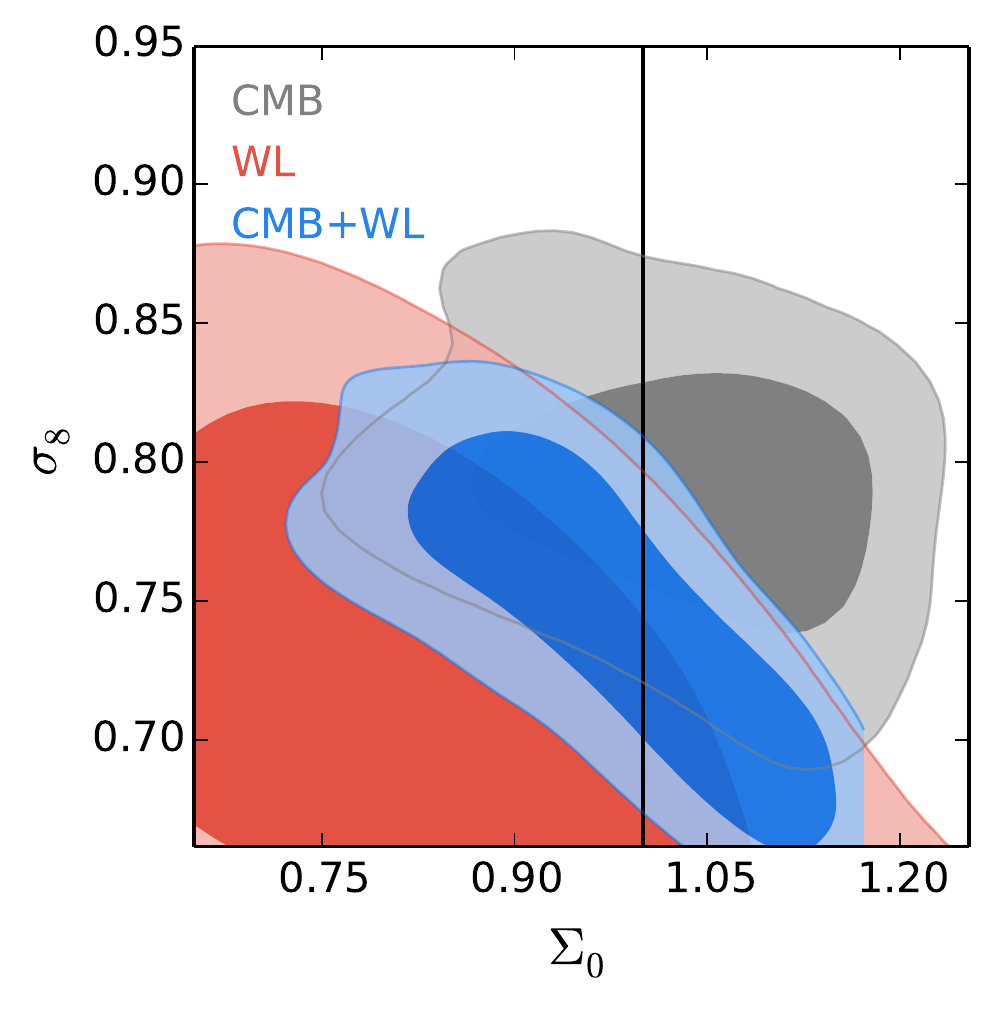}} &
{\includegraphics[width=2.2in,height=1.65in,angle=0]{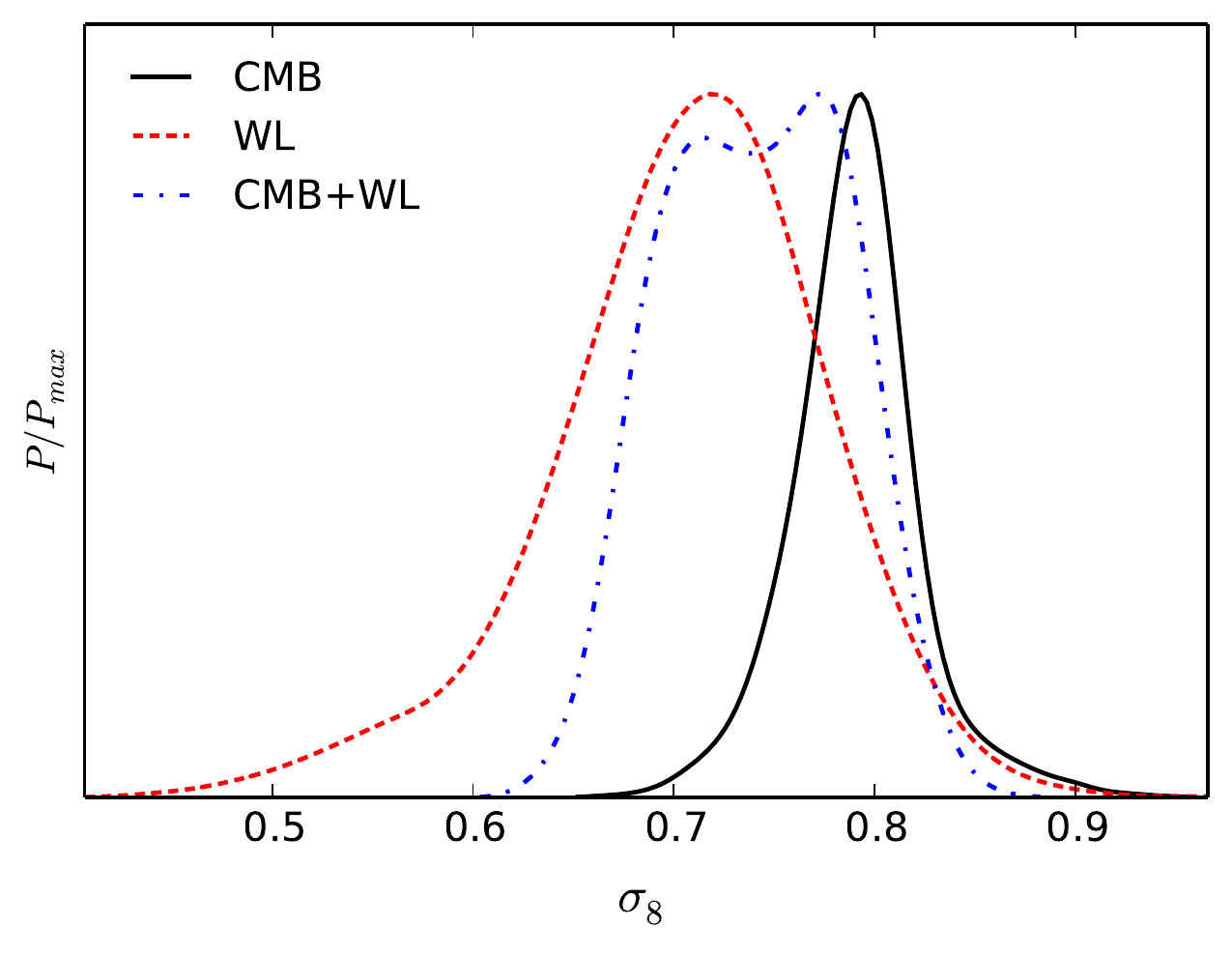}}&
{\includegraphics[width=2.2in,height=1.65in,angle=0]{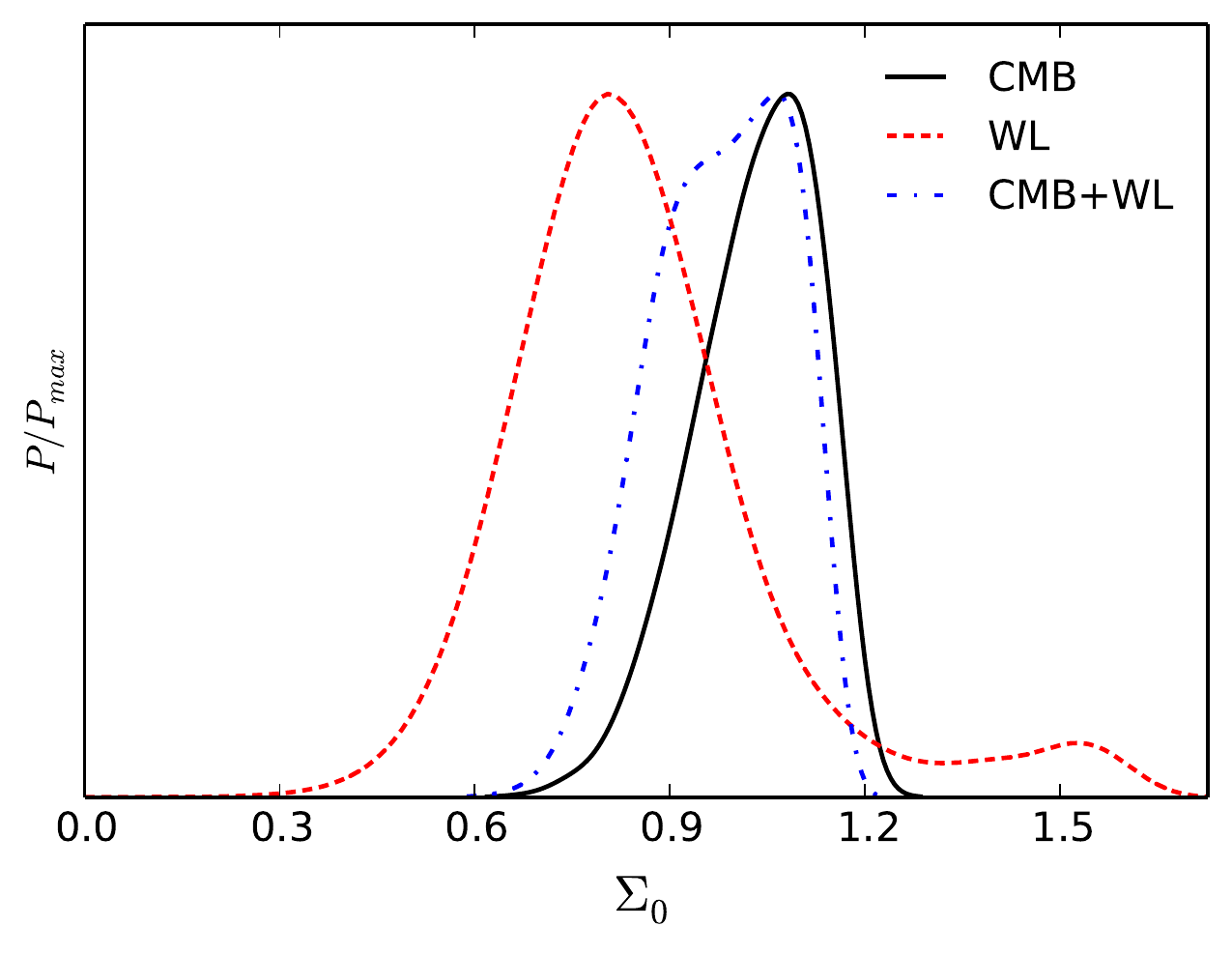}}
\end{tabular}
\caption{\label{fig:FUNCs8}
To illustrate that the tensions in the MG parameters may be arising from tensions between the Planck and CFHTLenS data sets, we plot the $68\%$ and $95\%$ 2-D confidence contours between $\Sigma_0$ and $\sigma_8$ as well as the 1-D probability distribution for $\sigma_8$  and $\Sigma_0$ when using method \textbf{P3} for evolving the MG parameters. We plot three different data set combinations: {\bf CMB} which includes all the data except CFHTLens; {\bf WL} which includes CFHTLenS but not the CMB data; and {\bf CMB + WL} which is the combination of all the data.  The probability distribution for $\sigma_8$ is significantly bimodal with peaks corresponding to the preferred values of $\sigma_8$ for the Planck and CFHTLenS data sets. From the plots of the 2-D confidence contours, one can see that the bimodality of $\Sigma_0$ seen in Fig. \ref{fig:FUNC1D} is directly tied to that of $\sigma_8$ with the CFHTLenS data causing a preference for lower values of $\Sigma_0$ and CMB data from Planck having a preference for higher values of $\Sigma_0$.} 
\end{figure}

In Fig. \ref{fig:FUNCs8} we show the 2-D confidence contours in the $\Sigma_0,\sigma_8$ plane as well as the 1-D probability distribution of $\sigma_8$ and $\Sigma_0$ for different combinations of the data sets that highlight the tensions between CMB and weak lensing (WL).  The first thing that can be noticed is $\sigma_8$, even more so than $\Sigma_0$ is bimodal.  In fact the two peaks in the distribution correspond roughly to the preferred $\sigma_8$ values of the CMB and weak lensing data sets.  This is because, compared to GR, the MG parameters allow the CMB to fit the Planck data better with lower values of $\sigma_8$ and at the same time allow for a better fit to the the weak lensing data with higher values of $\sigma_8$.  Quite importantly, from the 2-D confidence contours we see the higher values of $\Sigma_0$ occur in the region of the parameter space where the lower $\sigma_8$ values are preferred.   This lends credence to the conclusion that the higher preferred $\Sigma_1$ from \textbf{P2} is also being caused by this tension between the CMB and weak lensing data sets.

While the tensions with GR we have seen in this work are still weak, they are nonetheless more significant than we have seen in the past.  The fact that it is primarily due to a tension between two cosmological observations probing different eras of cosmic evolution is also important. Other explanations for the tensions between these data sets such as changes to the neutrino sector have been proposed, though some arguments have been made that these do not completely explain the tensions seen \cite{MacCrann:2014}.  It remains an open question whether these observed tensions are signaling some issues with the observations or underlying theory.  

\subsection{Galaxy Intrinsic Alignments: Constraints and Correlations}
\begin{figure}[h!]
\centering
\begin{tabular}{cm{1.95in}m{1.95in}}

\textbf{GR} & {\includegraphics[width=1.95in,height=1.95in,angle=0]{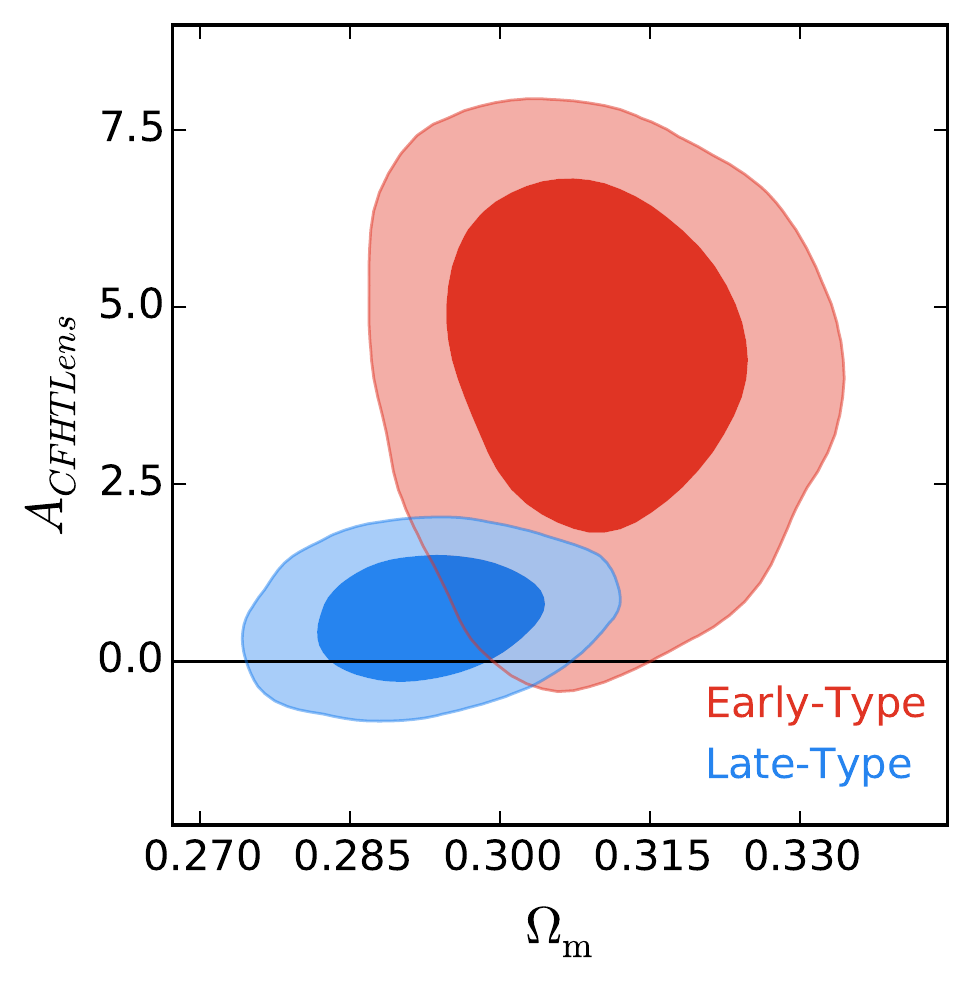}} &
{\includegraphics[width=1.95in,height=1.95in,angle=0]{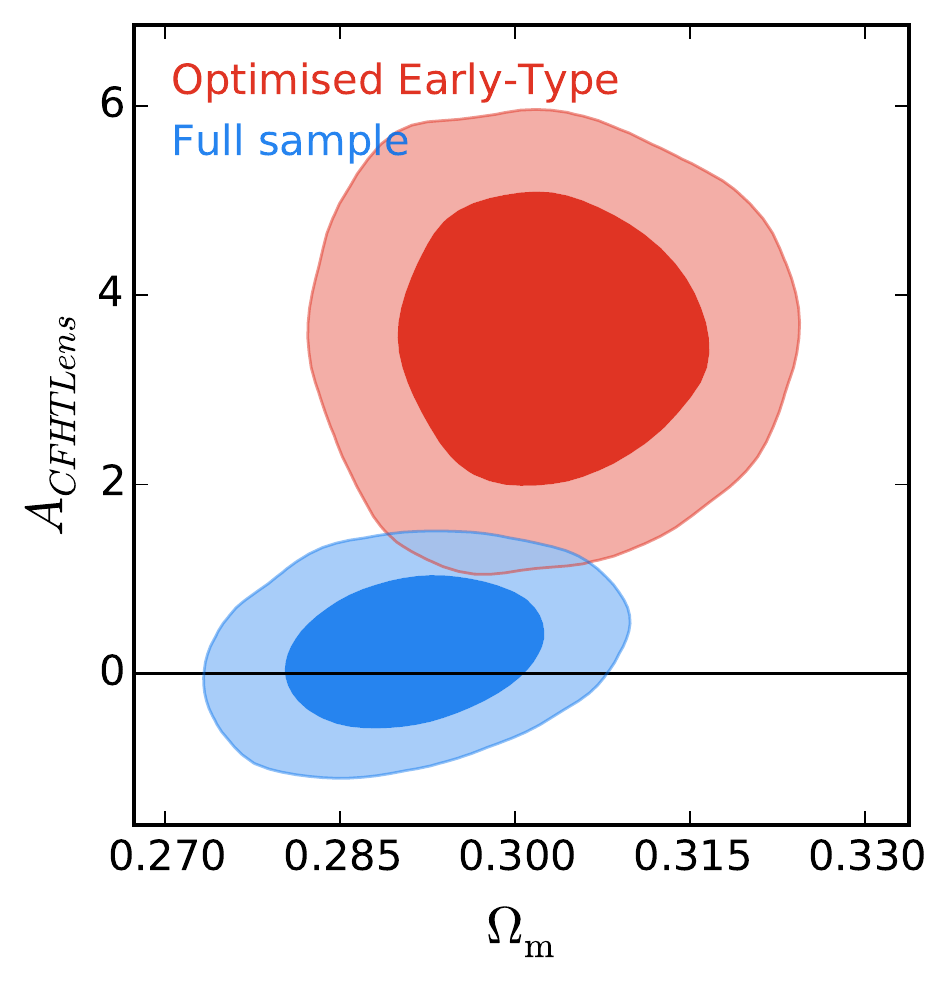}} \\
\textbf{P1} & {\includegraphics[width=1.95in,height=1.95in,angle=0]{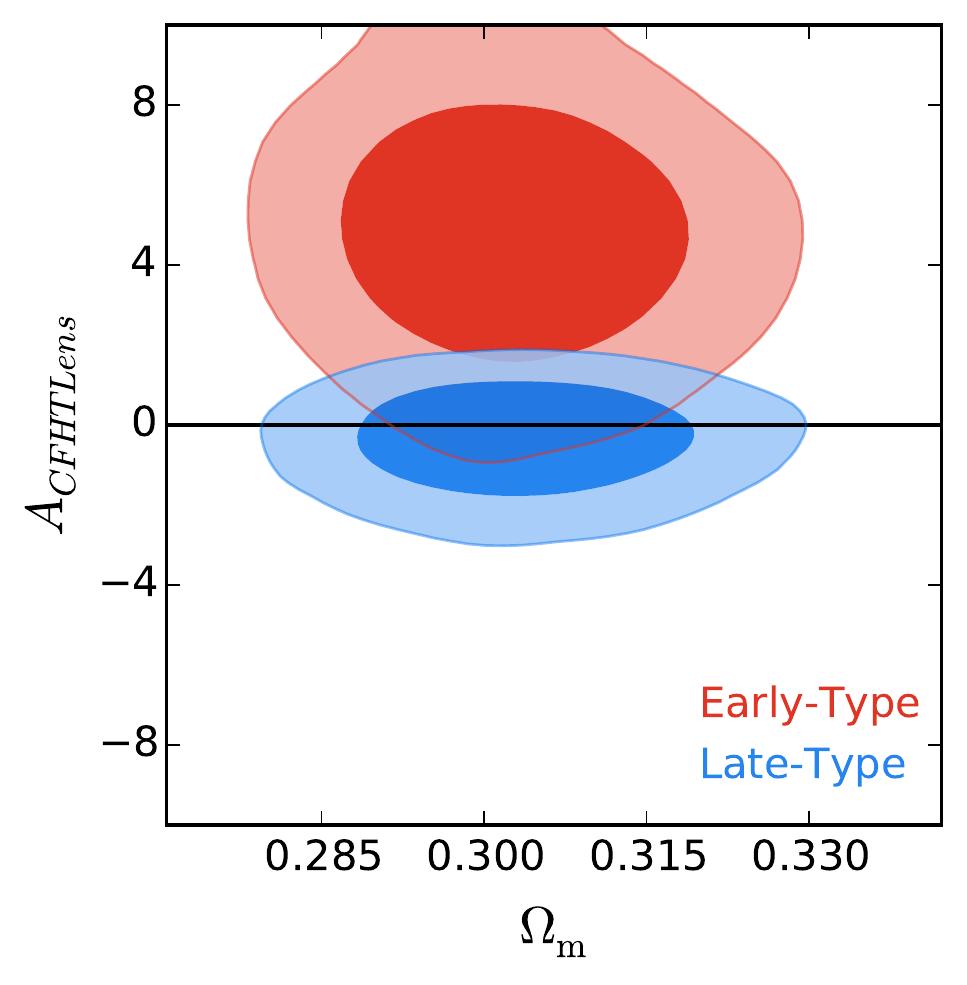}} &
{\includegraphics[width=1.95in,height=1.95in,angle=0]{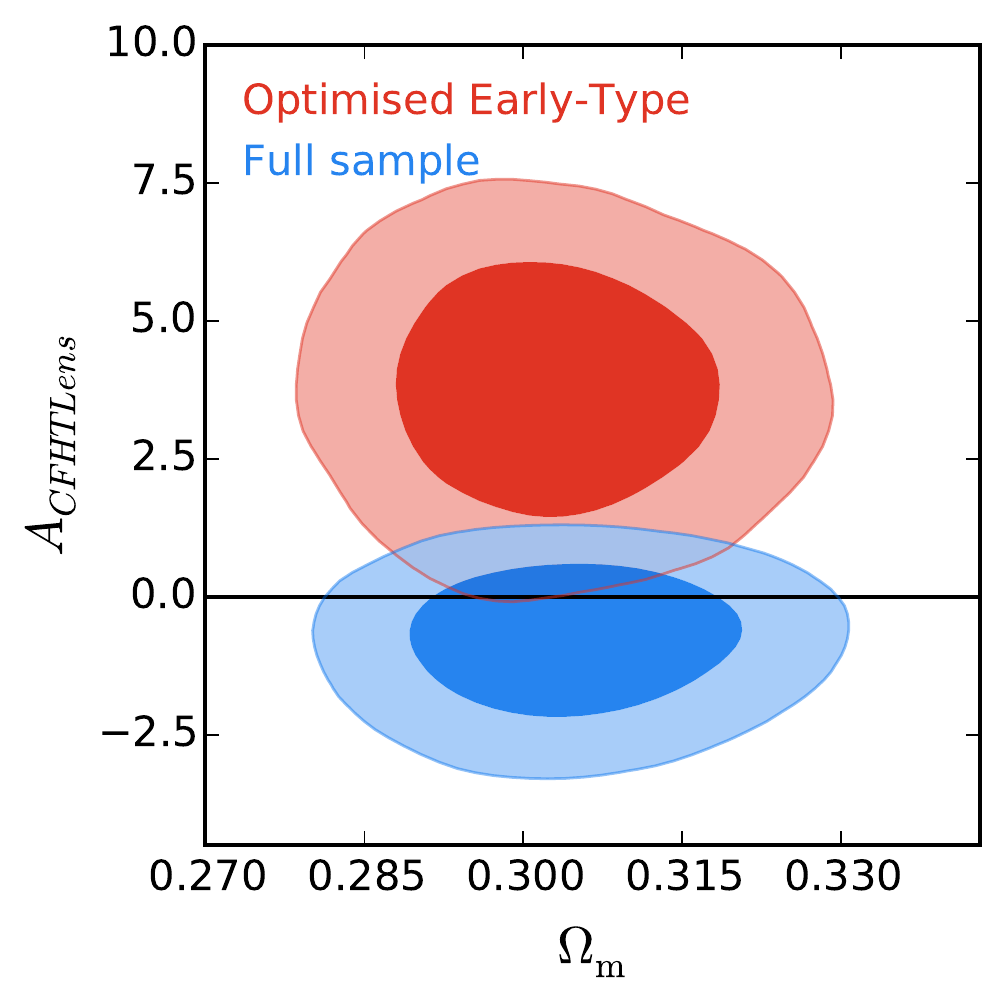}} \\
\textbf{P2} & {\includegraphics[width=1.95in,height=1.95in,angle=0]{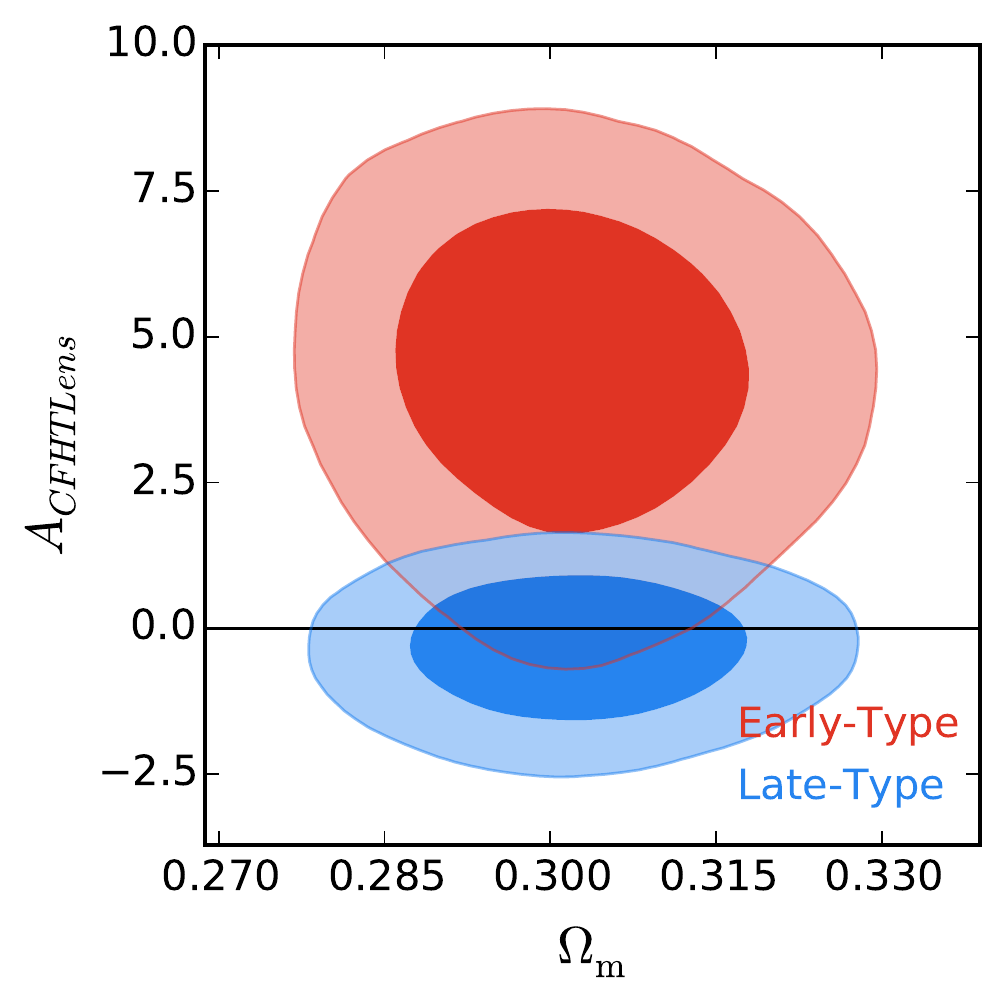}} &
{\includegraphics[width=1.95in,height=1.95in,angle=0]{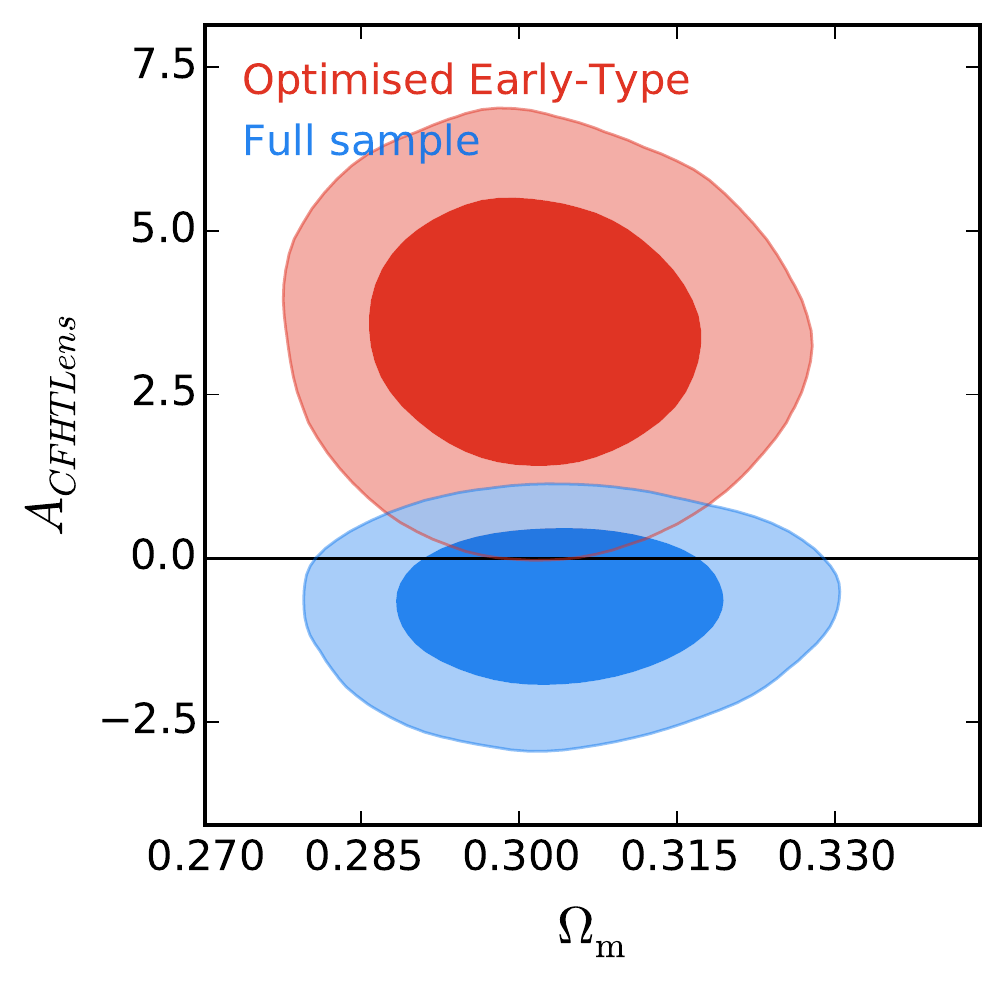}} \\
\textbf{P3} & {\includegraphics[width=1.95in,height=1.95in,angle=0]{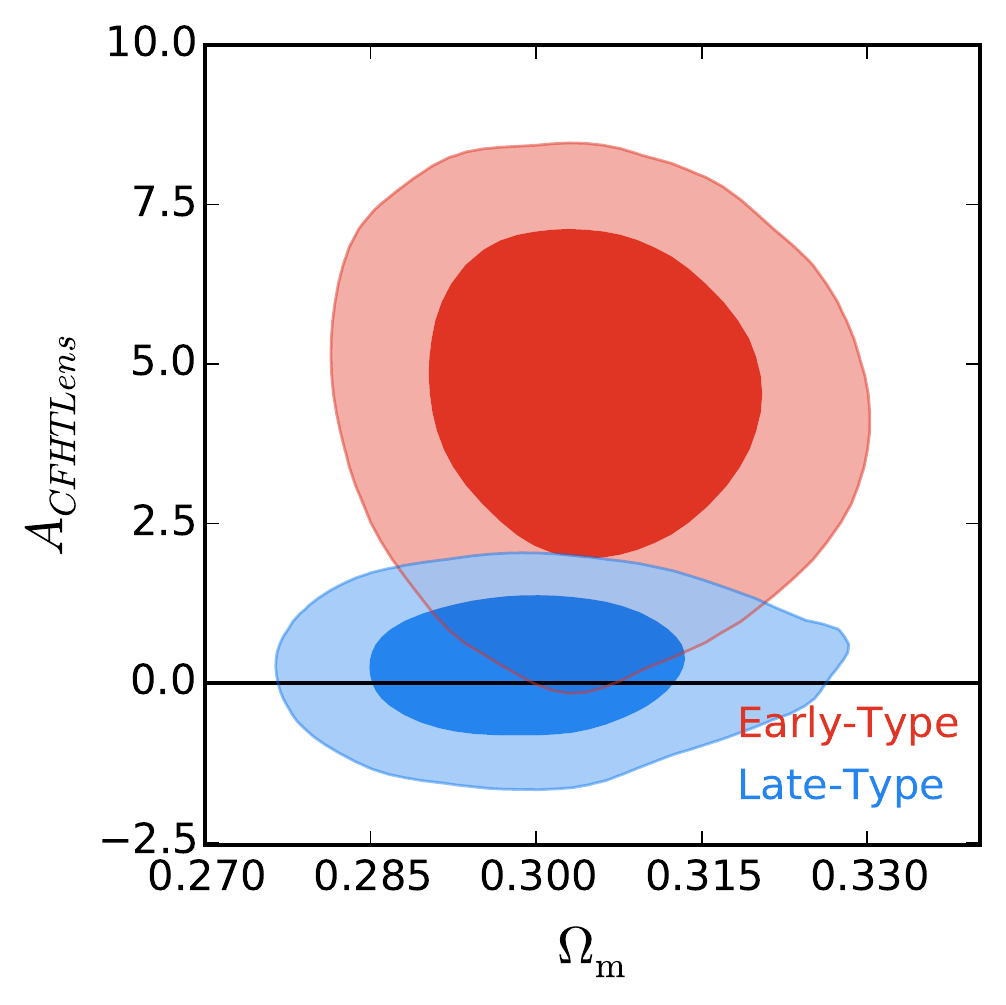}} &
{\includegraphics[width=1.95in,height=1.95in,angle=0]{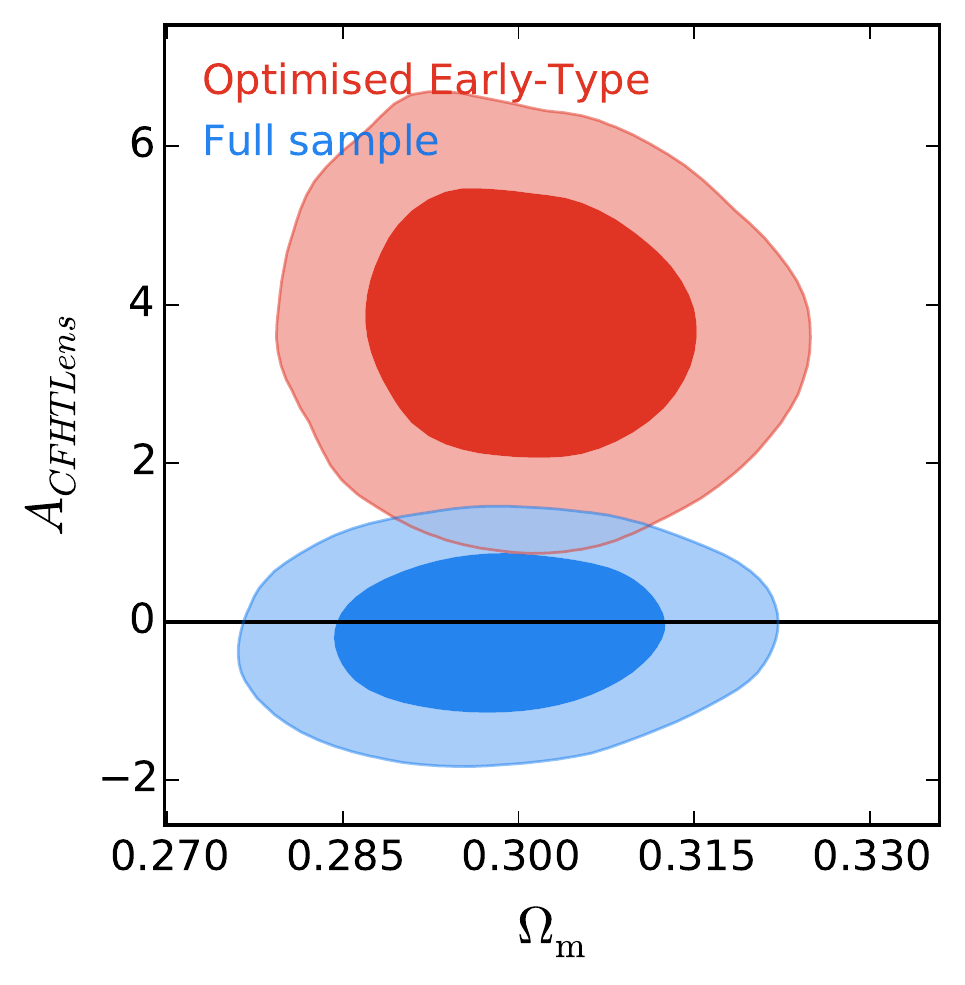}} \\

\end{tabular}
\caption{\label{fig:IA1}$68\%$ and $95\%$ 2-D confidence contours for the intrinsic alignment amplitude parameter $A_{\rm CFHTLenS}$ and $\Omega_m$. FIRST ROW: the theory is fixed to GR and the constraints obtained are in good agreement with those of \cite{Heymans:2013} though improved due to more precise recent data. To the left are the results for the intrinsic alignment optimized red galaxy sample of \cite{Heymans:2013}. The clear detection of non-zero $A_{\rm CFHTLenS}$ is also in agreement with \cite{Heymans:2013}. SECOND and THIRD ROWS: Similar constraints are presented but for the scale-dependent parameterizations \textbf{P2} and \textbf{P3} that model any deviation from GR. The bounds are larger but a zero $A_{\rm CFHTLenS}$ parameter is practically on the the $95\%$ CL boundary line for the optimized red galaxy sample.   
FOURTH ROW: results for the scale independent MG parameterization. The constraints are very similar to the GR case with a robust non-zero $A_{\rm CFHTLenS}$. This may hint to the effect of scale dependence in the MG parameterizations}
\end{figure}
\begin{figure}[h!]
\centering
\begin{tabular}{ccc}
{\includegraphics[width=2.in,height=2.in,angle=0]{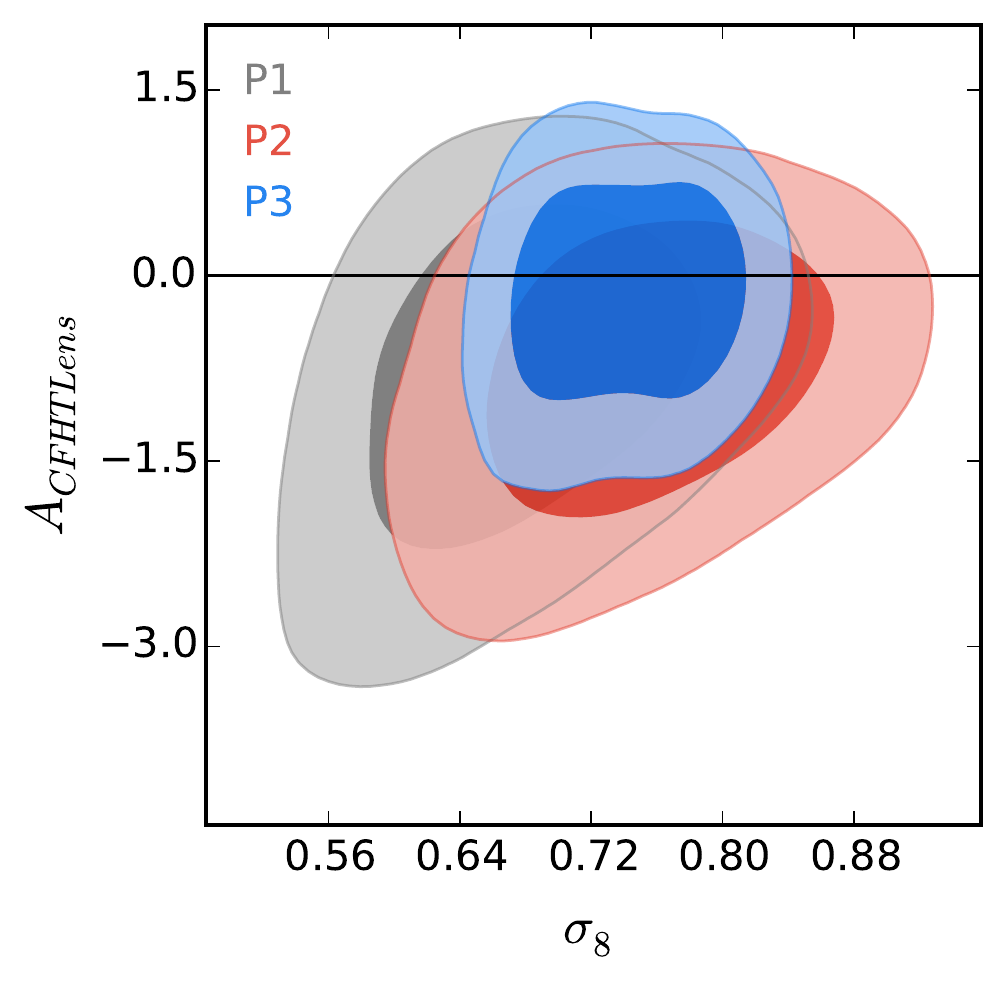}} &
{\includegraphics[width=2.in,height=2.in,angle=0]{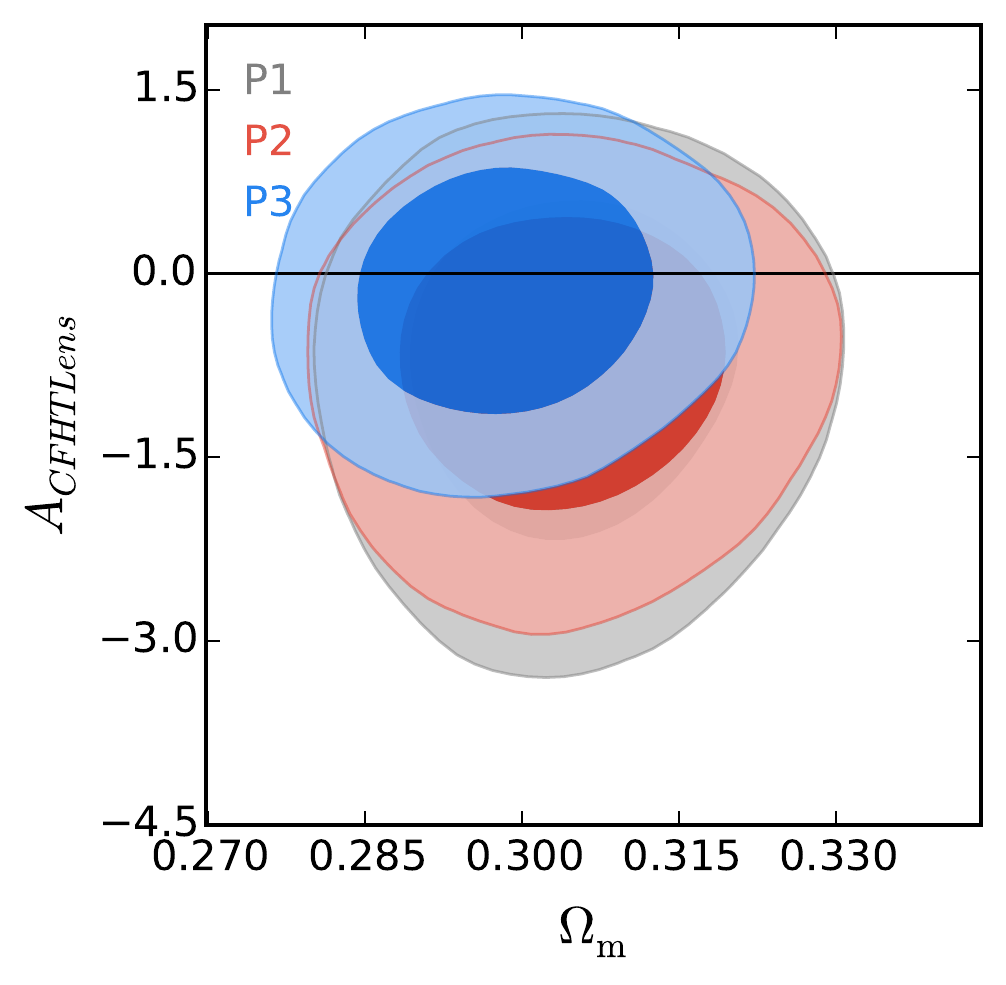}}&
{\includegraphics[width=2.7in,height=2.in,angle=0]{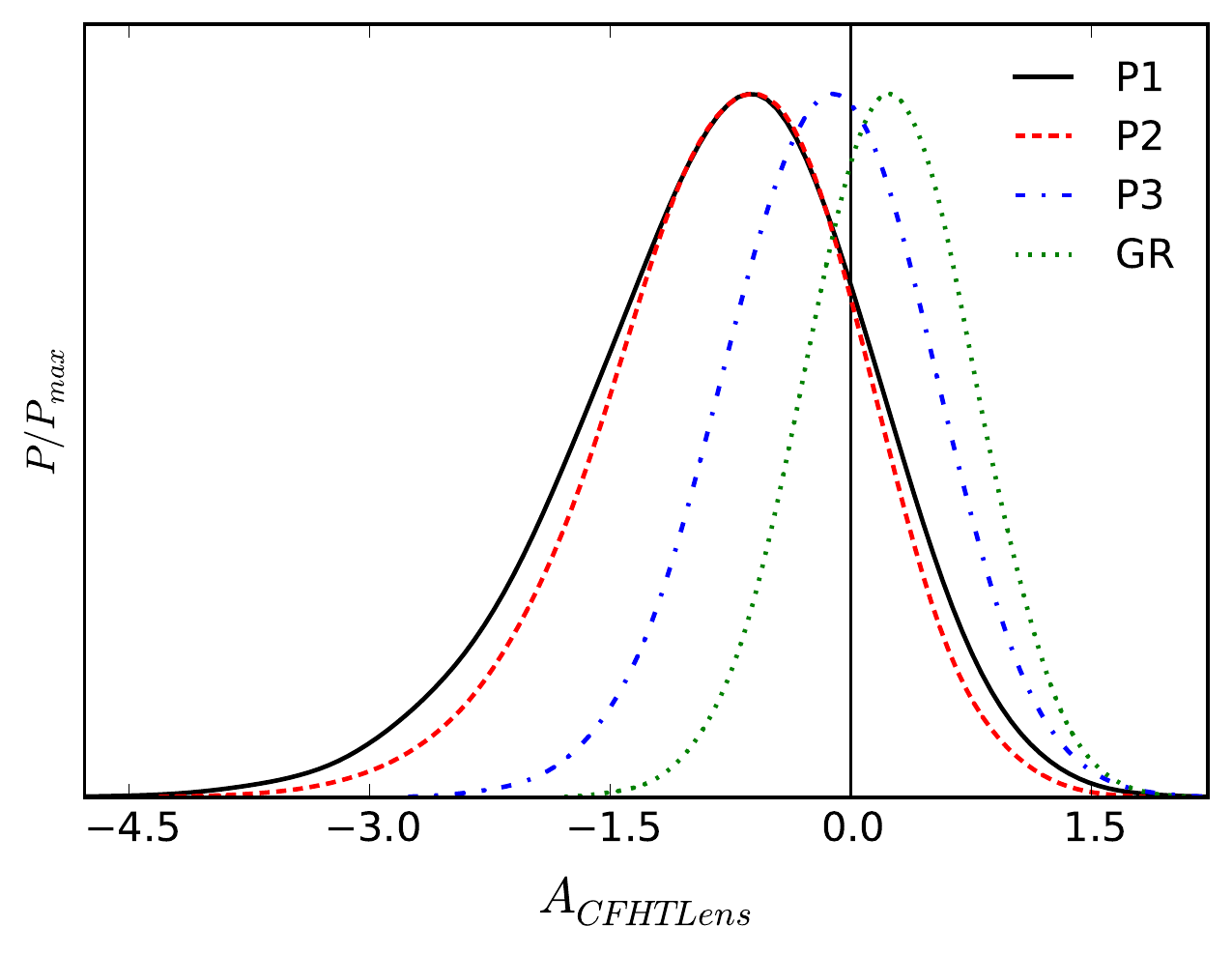}} \\
{\includegraphics[width=2.in,height=2.in,angle=0]{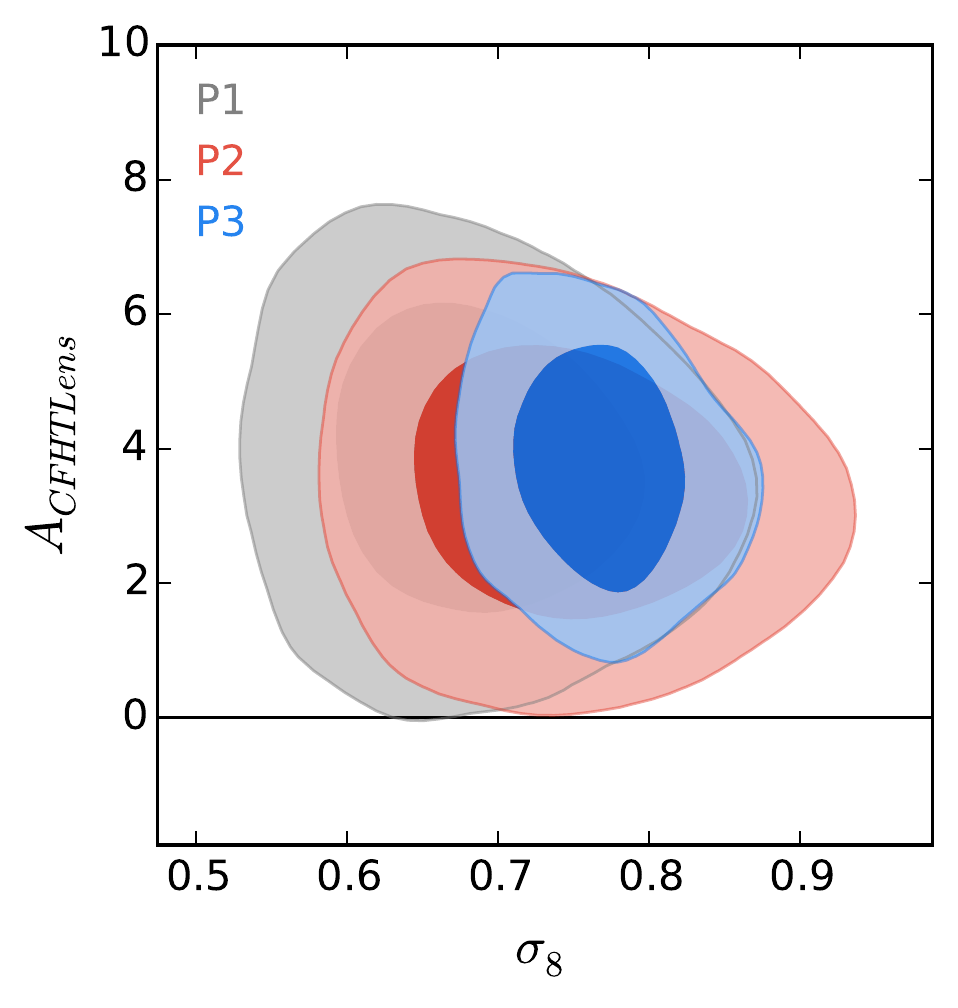}} &
{\includegraphics[width=2.in,height=2.in,angle=0]{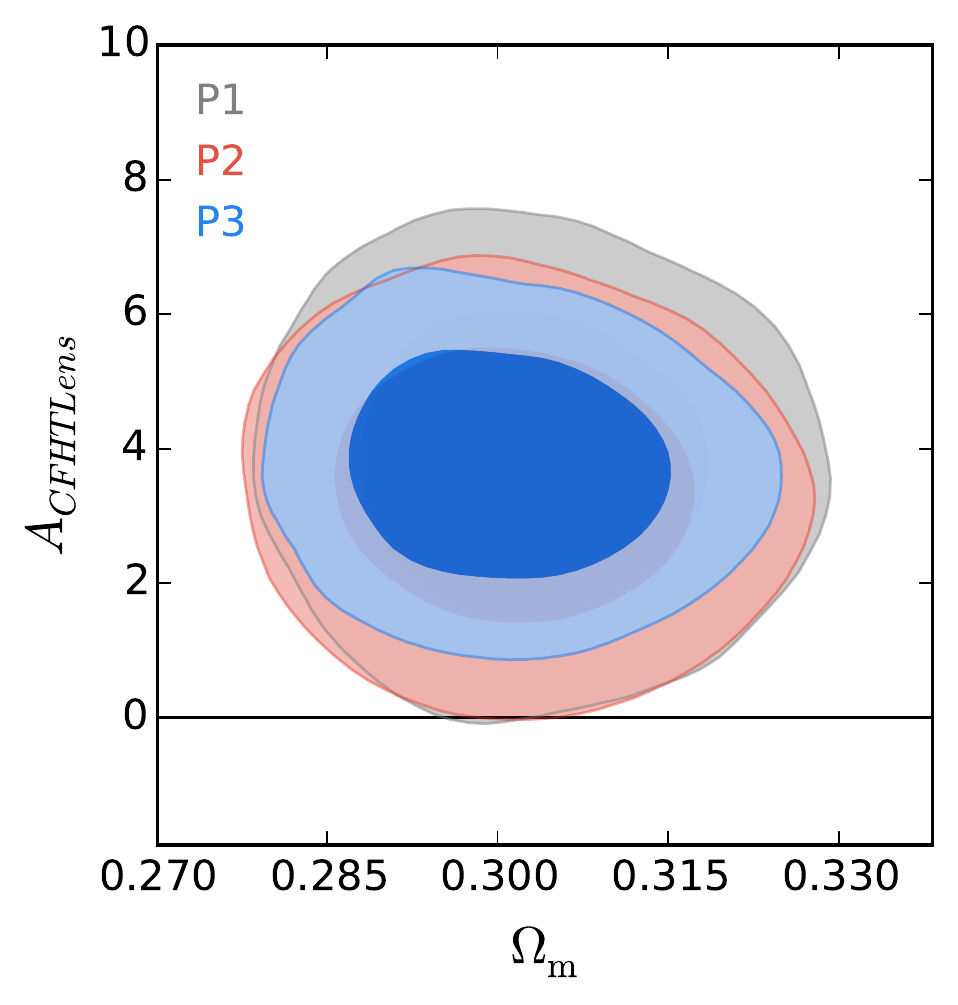}}&
{\includegraphics[width=2.7in,height=2.in,angle=0]{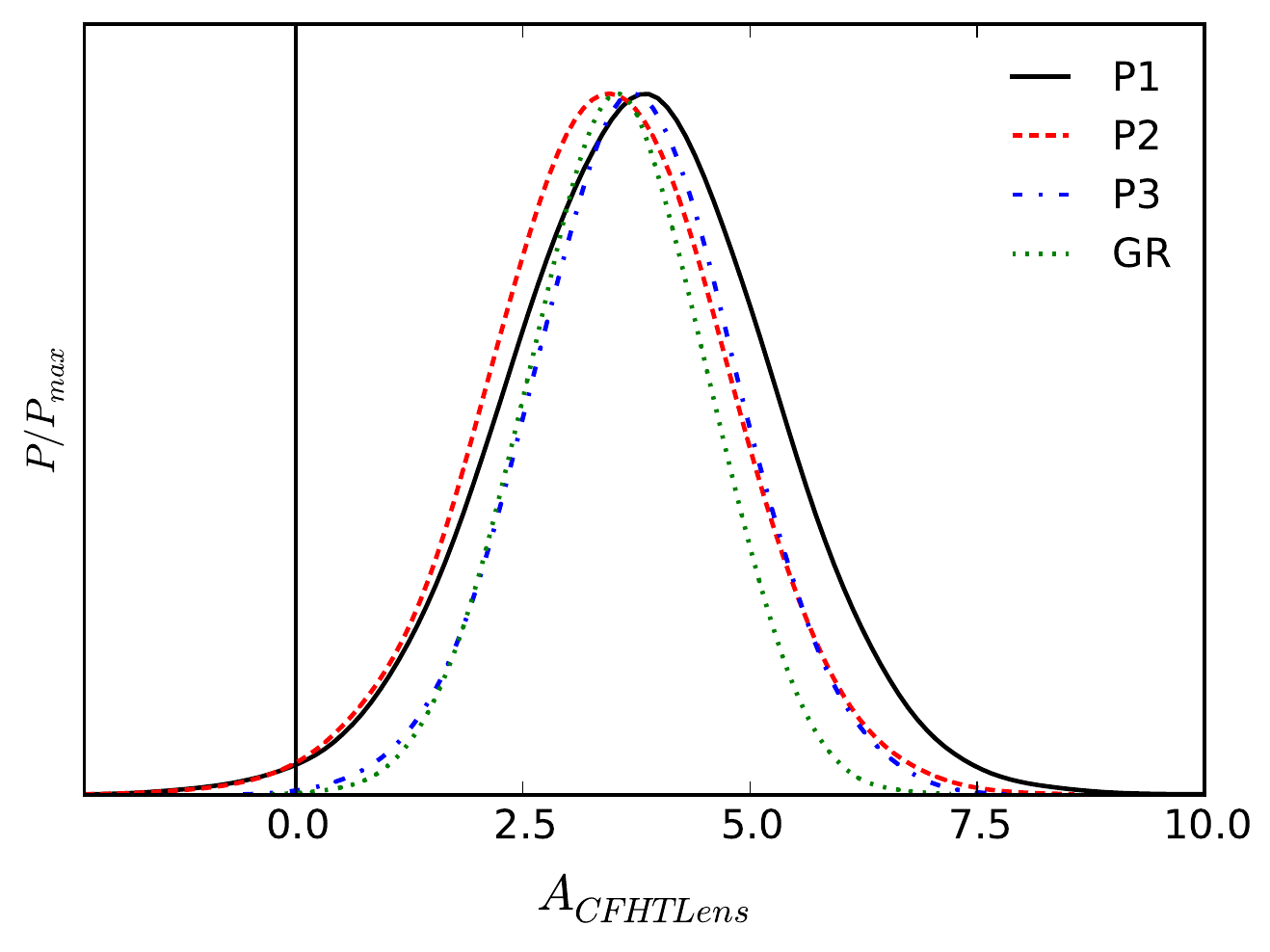}} \\
\end{tabular}
\caption{\label{fig:IA2}LEFT and MIDDLE: comparative $68\%$ and $95\%$ 2-D confidence contours for the intrinsic alignment amplitude parameter $A_{\rm CFHTLenS}$ versus $\sigma_{8}$, and $A_{\rm CFHTLenS}$ versus $\Omega_m$, for the 3 MG parameterizations. RIGHT: 1-D probability distributions for the parameter $A_{\rm CFHTLenS}$.  TOP ROW: constraints for the full CFHTLenS galaxy sample. BOTTOM ROW: constraints for the optimized red-galaxy sample.}
\end{figure}
Our results are presented in Figs. \ref{fig:IA1}, \ref{fig:IA2}, Tab. \ref{table:IA1}, Tab. \ref{table:IA2} and Tab. \ref{table:IA3}. 
When the underlying theory is fixed to GR, we find constraints on the 
 the amplitude of the intrinsic alignment model, $A_{\rm CFHTLenS}$ that are consistent with zero for the CFHTLenS full galaxy sample, the blue or the red samples (Figs. \ref{fig:IA1}, \ref{fig:IA2}). But we find a clear non-zero  $A_{\rm CFHTLenS}$ parameter when we use the optimized-red galaxy sample of \cite{Heymans:2013}, in agreement with their results. This is expected since the optimized sample uses red-foreground and blue-background galaxies thus maximizing the signal to noise of the GI measurement for the red galaxies, as found. For GR, we find for this sample $A_{\rm CFHTLenS}= 3.54 \pm 0.98$ for the $68\%$ limits. 
We find in general similar results to those of GR when the scale independent MG parameterization, \textbf{P3}, is used.  
However, when we use the binned and scale dependent MG parameterizations, \textbf{P1} and \textbf{P2}, we find that the zero $A_{\rm CFHTLenS}$ parameter is on the boundary line of the $95\%$ confidence contours.  
This is possibly due to the larger parameter space in these two cases leading to larger coutours. 
From the correlations tables, there are overall only weak to moderate correlations between the MG parameters and the intrinsic alignment amplitude parameter. It is found that $Q_2$ and $\Sigma_2$ (i.e. smaller scale and lower redshift bin) are the MG parameters most correlated with $A_{\rm CFHTLenS}$. This is the bin where most of the CFHTLenS data resides as can be seen in figure 1 of \cite{Heymans:2013}. We find no bimodality in the parameter $A_{\rm CFHTLenS}$ but different distributions depending on the MG parameterization with \textbf{P3} being the closest to that of GR. We also find only a moderate relative change in the figure of merit (i.e. $1.3-5.3\%$) for the MG parameters when $A_{\rm CFHTLenS}$ is fixed versus when it is varied showing a moderate  effect but this is likely to change for future high precision survey. 
%
\section{Conclusion}

In this work we have placed constraints on deviations from general relativity using the modified growth formalism and a combination of the latest cosmological data sets including: observations of the CMB anisotropy power spectrum from  the Planck satellite; observations of the galaxy power spectrum from the WiggleZ Dark Energy Survey; weak lensing tomography shear-shear cross correlations from the CFHTLenS survey; ISW-galaxy cross correlations; and BAO observations from 6dF, SDSS DR7, and BOSS DR9. We have also included in the analysis the effect of galaxy intrinsic alignment as a systematic effect in the weak lensing data. We used a model with a single nuisance parameter $A_{\rm CFHTLens}$ to account for intrinsic alignments. We expressed the results on modified gravity in terms of the parameters $Q$ and $\Sigma$.

The constraints obtained using these latest data sets give a $40-53\%$ improvement on the figure of merit for the MG parameters compared to previous studies \cite{Dossett:2011b}. GR is found to be consistent with observations according to all MG parameters when the 2-D marginalized 95\% confidence contours are considered.
\begin{table}[h!]		
\centering					
\begin{tabular}{|V|c|c|}\hline 								
\multicolumn{3}{|c|}{Correlation table}\\
\multicolumn{3}{|c|}{{Binning parameterization (\textbf{P1})}}\\	
\hline								
	&	$\sigma_8$	&	$\Omega_m$	\\ \hline			
$A_{\rm CFHTLenS}$	&	0.38292	&	0.052433	\\ \hline			
\multicolumn{3}{|c|}{{Hybrid parameterization (\textbf{P2})}}\\					\hline			
$A_{\rm CFHTLenS}$	&	0.32715	&	0.046858	\\ \hline			
\multicolumn{3}{|c|}{{Functional parameterization (\textbf{P3})}}\\					\hline
$A_{\rm CFHTLenS}$	&	0.12834	&	0.089536	\\ \hline			
\end{tabular}								
\caption{\label{table:IA3}Correlations between the intrinsic alignment amplitude parameter $A_{\rm CFHTLenS}$ and the amplitude of matter fluctuations $\sigma_8$ as well as the matter density parameter $\Omega_m$, for the three MG parameterizations.}			
\end{table}		 
We derived bounds and correlations of amplitude of the intrinsic alignment, $A_{\rm CFHTLenS}$. We included results for CFHTLenS galaxies split into early-type (red) and late-type (blue), and the optimized early-type sample (red-foreground-blue-background to maximize GI intrinsic alignments) of \cite{Heymans:2013}.    
The amplitude is found consistent with zero for the whole galaxy sample when the theory  is fixed to GR as well as when modified gravity is allowed. But a significantly ($95\%$ CL) non-zero $A_{\rm CFHTLenS}$ is found when the optimized early-type sample is used for GR (i.e. $A_{\rm CFHTLenS}= 3.54 \pm 0.98$ for the $68\%$ CL, consistent and slightly improved limits with \cite{Heymans:2013}). We obtain similar detection results when our functional scale-independent MG parameterization \textbf{P3} is used. For our binned scale-dependent MG parameterizations (\textbf{P1} and \textbf{P2}), the bounds are larger and the zero $A_{\rm CFHTLenS}$ is on the border line of the $95\%$ confidence contours, most likely due to a larger parameter space in the binned cases. We find overall weak to moderate correlations between $A_{\rm CFHTLenS}$ and MG parameters, the largest ones being with our parameters $Q_2$ and $\Sigma_2$.
Only a relatively moderate change in the figure of merit (i.e. $1.25-5.30\%$) for the MG parameters is found when we include or not intrinsic alignments in the analysis but the effect is likely to be more pronounced for future high precision survey. 

For our MG parameterization \textbf{P3}, we obtained a bimodal probability distribution for the MG parameter $\Sigma_0$ indicating the presence of some tensions between the data sets.
We analyzed some aspects of these tensions and find them related or similar to what is observed with the amplitude of matter fluctuation parameter $\sigma_8$. Indeed, we find that the data seems to prefer a lower value of the modified gravity parameter $\Sigma_0$ in parts of the parameter space where higher values of $\sigma_8$ are preferred, while areas of the parameter space that prefer a lower value of $\sigma_8$  favor a higher value of $\Sigma_0$.  This finding is consistent with tensions in the MG parameter space from previous studies and also with tensions in $\sigma_8$ reported in other recent works. Further work is needed in order to investigate the source of these tensions with the incoming higher precision data and surveys. 
\acknowledgments
We thank C. Heymans for making the red-galaxy and the optimized red-galaxy samples available via the CFHTLenS website. We thank A. Peel for reading the manuscript and M. Troxel for useful comments. J.D acknowledges support from the European Research Council through the Darklight ERC Advanced Research Grant (\# 291521).  M.I. acknowledges that this material is based upon work supported in part by NSF grant AST-1109667, and that part of the calculations for this work have been performed on the Cosmology Computer Cluster at UT-Dallas funded by the Hoblitzelle Foundation. D.P. was supported by an Australian Research Council Future Fellowship [grant number FT130101086]. Parts of this research were conducted by the Australian Research Council Centre of Excellence for All-sky Astrophysics (CAASTRO), through project number CE110001020.

\end{document}